\documentclass[useAMS,usenatbib]{mn2e}
\usepackage[]{hyperref}
\usepackage{color}
\usepackage[utf8]{inputenc}
\usepackage{graphicx,url,times}
\newif\ifpdf
  \ifx\pdfoutput\undefined
    \pdffalse
  \else
    \ifnum\pdfoutput=1
      \pdftrue
    \else
      \pdffalse
    \fi
  \fi

\ifpdf
    \graphicspath{{plots/}}
\else
    \graphicspath{{eps/}}
\fi

\newcommand{\papertitle}{Subhaloes gone Notts: Spin across subhaloes and finders}
\ifpdf\hypersetup{
pdftitle={\papertitle},
pdfauthor={Julian Onions},
pdfkeywords={N-body simulations, haloes evolution, dark matter},
pdfstartview=FitH,
}
\fi

\newcommand{\ahf}{\textsc{ahf}}
\newcommand{\subfind}{\textsc{subfind}}

\newcommand{\gadget}{\textsc{gadget3}}
\newcommand{\voboz}{\textsc{voboz}}
\newcommand{\mendieta}{\textsc{mendieta}}

\newcommand{\adaptahop}{\textsc{adaptahop}}

\newcommand{\htd}{\textsc{hot3d}}
\newcommand{\hsd}{\textsc{hot6d}}

\newcommand{\rockstar}{\textsc{rockstar}}
\newcommand{\hbt}{\textsc{hbt}}
\newcommand{\ghp}{\textsc{grasshopper}}
\newcommand{\skid}{\textsc{skid}}

\newcommand{\stf}{\textsc{stf}}
\newcommand{\hkpc}{{\ifmmode{h^{-1}{\rm kpc}}\else{$h^{-1}$kpc}\fi}}
\newcommand{\abs}[1]{\left|#1\right|}
\newcommand{\ghalo}{\textsc{ghalo}}
\newcommand{\Fig}[1]{Figure~\ref{#1}}
\newcommand{\Eqn}[1]{Equation~\ref{#1}}
\newcommand{\Tbl}[1]{Table~\ref{#1}}
\def\vmax{$v_{\rm max}$}
\def\Rmax{$r_{\rm max}$}

\newcommand{\subhalos}{subhaloes}
\newcommand{\Sec}[1]{Section~\ref{#1}}
\newcommand{\Ssec}[1]{subsection~\ref{#1}}

\newlength{\figwidth}
\newlength{\resplot}
\title {\papertitle}
\author[Onions et al.]
{Julian~Onions,$^1$\thanks{E-mail: \href{mailto:julian.onions@gmail.com}{julian.onions@gmail.com}}
  Yago Ascasibar,$^2$
  Peter Behroozi,$^{3,4,5}$
  Javier~Casado,$^2$
\newauthor
  Pascal Elahi,$^{6,1}$
  Jiaxin Han,$^{6,7,8}$
  Alexander~Knebe,$^2$
  Hanni~Lux,$^1$
\newauthor
  Manuel E. Merch\'{a}n,$^{9}$
  Stuart~I.~Muldrew,$^1$ 
  Mark Neyrinck,$^{10}$
  Lyndsay~Old,$^1$ 
\newauthor
  Frazer~R.~Pearce,$^1$ 
  Doug~Potter,$^{11}$
  Andr\'{e}s N. Ruiz,$^{9}$ 
  Mario A. Sgr\'{o},$^{9}$ 
\newauthor
  Dylan Tweed$^{12}$
  and Thomas~Yue$^1$ 
\\
  $^1$School of Physics \& Astronomy, University of Nottingham, Nottingham, NG7 2RD, UK\\
  $^2$Departamento de F\'{ı}sica Te\'{o}rica, M\'{o}dulo C-15, Facultad de Ciencias, 
  Universidad Aut\'{o}noma de Madrid, 28049 Cantoblanco, Madrid, Spain\\
  $^{3}$Kavli Institute for Particle Astrophysics and Cosmology, Stanford, CA 94309, USA\\
  $^{4}$Physics Department, Stanford University, Stanford, CA 94305, USA\\
  $^{5}$SLAC National Accelerator Laboratory, Menlo Park, CA 94025, USA\\
  $^{6}$Key Laboratory for Research in Galaxies and Cosmology, Shanghai Astronomical Observatory, Shanghai 200030, China\\
  $^{7}$Graduate School of the Chinese Academy of Sciences, 19A, Yuquan Road, Beijing, China \\
  $^{8}$Institute for Computational Cosmology, Department of Physics, Durham University, South Road, Durham DH1 3LE, UK\\
  $^{9}$Instituto de Astronom\'{ı}a Te\'{o}rica y Experimental (CCT C\'{o}rdoba, CONICET, UNC), Laprida 922, X5000BGT, C\'{o}rdoba, Argentina\\
  $^{10}$Department of Physics and Astronomy, Johns Hopkins University, 
  3701 San Martin Drive, Baltimore, MD 21218, USA\\
  $^{11}$Insitute for Theoretical Physics, Univ. of Z\"{u}rich, Winterthurerstrasse 190, CH-8057 Z\"{u}rich, Switzerland\\
  $^{12}$Racah Institute of Physics, The Hebrew University, Jerusalem 91904, Israel. \\
}

\begin{document}
\date{Accepted 2012 November 30. Received 2012 November 21; in original form 2012 August 28}

\pagerange{\pageref{firstpage}--\pageref{lastpage}} \pubyear{2012}\volume{0000}

\maketitle

\label{firstpage}

\begin{abstract}
  We present a study of a comparison of spin distributions of
  subhaloes found associated with a host halo. The subhaloes are found
  within two cosmological simulation families of Milky Way-like galaxies,
  namely the Aquarius and GHALO simulations.  These two simulations use
  different gravity codes and cosmologies. We employ ten different
  substructure finders, which span a wide range of methodologies
  from simple overdensity in configuration space to full 6-d phase
  space analysis of particles.  We subject the results to a common
  post-processing pipeline to analyse the results in a consistent
  manner, recovering the dimensionless spin parameter.  We find that
  spin distribution is an excellent indicator of how well the removal
  of background particles (unbinding) has been carried out.  We also
  find that the spin distribution decreases for substructure the nearer
  they are to the host halo's, and that the value of the spin parameter
  rises with enclosed mass towards the edge of the substructure. Finally
  subhaloes are less rotationally supported than field haloes, with the 
  peak of the spin distribution having a lower spin parameter.
\vspace{0.1cm} \end{abstract}


\section{Introduction} \label{sec:introduction}
Within the hierarchical galaxy formation model, dark matter haloes are
thought to play the role of gravitational building blocks, within
which baryonic diffuse matter collapses and becomes detectable
\citep{white_1978,white_1991}. Gravitational processes that determine
the abundance, the internal structure and kinematics, and the formation
paths of these dark haloes within the cosmological framework,
can be simulated in great detail using $N$-body methods. However, the
condensation of gas associated with these haloes, eventually leading
to stars and galaxies we see today, is still at the frontier of present
research efforts. A first exploration of the (cosmological) formation
of disc galaxies  has been presented in \citet{fall_1980}, where
it was shown that galactic spin is linked to the surrounding larger
scale structure (e.g. the parent halo). In particular, the general
theory put forward by Fall \& Efstathiou reproduces galactic discs with
roughly the right sizes, if specific angular momentum is conserved,
as baryons contract to form a disc (previously suggested by
\citet{mestel_1963}) and if baryons and dark matter initially share
the same distribution of specific angular momentum. 

While the theory has subsequently been refined, it
always included (and still includes) such a coupling between
the parent halo's angular momentum and the resulting galactic disc
(cf. \citet{dalcanton_1997, mo_1998, navarro_2000, abadi_2003,bett_2010}
).  The origin of the halo's spin can now be understood in terms of
tidal torque theory in which protohaloes gain angular momentum from
the surrounding shear field (e.g., \citet{peebles_1969, white_1984,
barnes_1987}) as well as by the build-up of angular momentum through
the cumulative transfer of angular momentum from subhalo accretion
\citep{vitvitska_2002}. Whichever way the halo gains its spin,
it is a crucial ingredient for galaxy formation and all semi-analytical
modelling of it \citep{kauffmann_1993,kauffmann_1997,frenk_1997,cole_2000,
benson_2001, croton_2006, delucia_2007, bower_2006, bertone_2007,
font_2008, benson_2012}.

A number of studies have been performed on the spin of haloes, in
particular studies by
\citet{peebles_1969,bullock_2001,hetznecker_2006,bett_2007,maccio_2007,
gottlober_2007,knebe_2008,antonuccio_2010,wang_2011b,
trowland_2012,lacerna_2012,bryan_2012} but so far little has been done
on subhaloes.  These studies look at the spin of individual dark
matter haloes found in cosmological simulations and generally do not
focus on the substructure, or differences between substructure
definition due to lack of resolution. Here we present a comparison of
spin parameters across a number of detected subhaloes found by a
variety of substructure finders. The finders use many different
techniques to detect substructure within a larger host halo. This is a
follow-up to a more general paper comparing the recovery of structure
by different finders in \citet{Onions_2012} and its predecessor
\citet{knebe_haloes_2011}.

The techniques studied here for finding substructures include real-space,
phase-space, velocity-space finders, as well as finders employing a
Voronoi tessellation, tracking haloes across time using snapshots,
friends-of-friends techniques, and refined meshes as the starting point
for locating substructure.  With such a variety of mechanisms and
algorithms, there is little chance of any systematic source of errors
in the collection of substructure distorting the result.  Subhaloes are
particularly subject to distortion and evolution, more so than haloes
because, by definition, they reside within a host halo with which they
tidally interact. This can affect their structure and other parameters,
and in this case we are particularly interested in the spin properties. We
quantify the spin with the parameter $\lambda$, a dimensionless quantity
that characterises the spin properties of a halo and is explained in
more detail in \Sec{sec:method}.

The rest of the paper is structured as follows. We first describe the
methods used to quantify the spin of the halo in \Sec{sec:method}.
The data we used is described in \Sec{sec:data}.  Next we look at the
overall properties of the spin in \Sec{sec:spin}.  Then we look at the
correlation between the host halo and the subhaloes spin in
\Ssec{ssec:radcomp}.  Finally we look at how the spin is built up
within the subhalo as a function of mass in \Ssec{ssec:submass}.  We
conclude in \Sec{sec:Summary}.

\section{Method}\label{sec:method}
\subsection{Spin parameter}
The dimensionless spin parameter gives an indication of how much a
gravitationally bound collection of particles is supported in equilibrium
via net rotation compared to its internal velocity dispersion.  The spin
parameter varies between 0, for a structure negligibly supported by
rotation, to values of order 1 where it is completely rotationally
supported, and in practice maximum values are usually $\lambda \approx
0.4$ \citep{frenk_2012}.  Values larger than 1 are unstable
structures not in equilibrium.

There are two variants of the spin parameter that are in common use.
\cite{peebles_1969} proposed to parametrise the spin using the expression
given in \Eqn{eqn:peebles}.
\begin{equation}
\lambda = \frac{J \sqrt{\abs{E}}}{GM^{5/2}}
 \label{eqn:peebles}
\end{equation}
where $J$ is total angular momentum, $E$ the energy and $M$ the mass of
the structure. In isolated haloes, all of these quantities are conserved,
which gives the definition a time independence.

\citet{bett_2007} measured the Peeble's spin parameter and fitted an
expression to the distribution for haloes extracted from the
Millennium simulation \citep{springel_millenium_2005};
that is characterised by \Eqn{eqn:bettfit}
\begin{equation}
P(\log \lambda) = A \left( \frac{\lambda}{\lambda_0} \right) ^3
\exp \left[ -\alpha \left(\frac{\lambda}{\lambda_0}\right)^{3/\alpha} \right]
 \label{eqn:bettfit}
\end{equation}
where A is 
\begin{equation}
 A = 3 \ln 10 \frac{\alpha^{\alpha-1}}{\Gamma(\alpha)}
\label{eqn:bettA}
\end{equation}
The variables $\lambda_0$ and $\alpha$ are free parameters, and $\Gamma(\alpha)$ is the gamma function.
The best fit they found for field haloes was with $\lambda_0 = 0.04326$ and $\alpha=2.509$.

\cite{bullock_2001} proposed a different definition of the spin
parameter, $\lambda'$, expressed in \Eqn{eqn:bullock}. As it is not
dependent on measuring the energy it is somewhat faster to calculate
when dealing with large numbers of haloes.

\begin{equation}
  \lambda' = \frac{J}{\sqrt{2}MRV}
 \label{eqn:bullock}
\end{equation}
Here $J$ is the angular momentum within the enclosing sphere of virial
radius $R$ and virial mass $M$, and $V$ is the circular velocity at the
virial radius ($V^2=GM/R$). The Bullock spin parameter is more robust
to the position of the outer radius of the structure. Bullock proposes a
fitting function to the distribution as described in \Eqn{eqn:bullockfit}
which was based on one from \citet{barnes_1987}.

\begin{equation}
 P(\lambda') = \frac{1}{\lambda' \sqrt{2\pi} \sigma} 
\exp \left( - \frac{ \ln^2 (\lambda' / \lambda_0')}{ {2\sigma^2}}  \right)
\label{eqn:bullockfit}
\end{equation}
This has free parameters $\lambda_0'$ and $\sigma$ and \citet{bullock_2001}
found a best fit for field haloes at values of
$\lambda_0' = 0.035$ and $\sigma = 0.5$.

The Peebles calculation is perhaps more well defined for a given set
of particles, as it is calculated directly from the particles
properties, whereas the Bullock parameter is easier to calculate from
gross halo statistics, and is not dependant on the density
profile. For more comparisons of the two parameters the reader is 
referred to \citet{hetznecker_2006}

\subsection{The SubHalo Finders} \label{ssec:Finders}

In this section we briefly list the halo finders that took part in the
comparison project. More details about the specific algorithms are
available in \citet{Onions_2012} and the articles referenced therein.

\begin{itemize} 
\item \adaptahop\ (Tweed) is a configuration space over density finder \citep{adaptahop_2004,adaptahop_2009}.

\item \ahf\ (Knollmann \& Knebe) is a configuration space spherical overdensity adaptive 
mesh finder \citep{gill_evolution_2004,knollmann_ahf:_2009}.

\item 
\ghp\ (GRadient ASSisted HOP) (Stadel) is a reworking of the \skid\ group
finder\citep{stadel_2001} and appears within our wider comparison for
the first time here, and so is described in more detail. It takes an
approach like the HOP algorithm \citep{hop_1998} and reproduces the
initial grouping of \skid\ in two computational steps. First densities are
calculated for all particles as before using the Monaghan M3 SPH kernel
over 80 nearest neighbours. Second, for each particle, the gradient of the
density is calculated in a way that cancels the so-called E0 error in the
gradient \citep{read_2010}, by using the gradient of the M3 kernel. Then,
given that the neighbours lie within a ball of radius $2h$, we create a
link from this particle to the closest neighbour to the point a distance
$h$ in the direction of the gradient.

After links have been created, each particle follows the chain
of links until is reaches a cycle, marking oscillation about the density
peak of the group. Finally since noise below a gravitational softening
length causes a lot of artificial density peaks we search for particles
of a cycle which are within a distance $\tau$ of any other particles in a
cycle. The parameter $\tau$ is typically set to 4 times the gravitational
softening length, as was the typical case for \skid. Unbinding is also
performed in a nearly equivalent way to \skid, but now scales as $O(n
\log n)$ as opposed to $O(n^2)$ as was the case with the original \skid.

The group finding with \ghp\ is now fast enough to allow it to be
performed during a simulation but gives nearly identical results to the
previous \skid\ algorithm.

\item Hierarchical Bound-Tracing (\hbt) (Han) is a tracking algorithm
working in the time domain \citep{han_resolving_2011}.

\item HOT+FiEstAS (\htd\ \& \hsd) (Ascasibar) is a general-purpose 
clustering analysis tool, working either in configuration or phase space
\citep{ascasibar_2005,ascasibar_2010}.

\item \mendieta\ (Sgr\'{o}, Ruiz \& Merch\'an) is a Friends-of-Friends 
based finder that works in configuration space \citep{sgro_2010}.

\item \rockstar\ (Behroozi) is a phase-space halo finder \citep{behroozi_2011}.  

\item \stf\ (Elahi) is a velocity space/phase-space finder \citep{elahi_peaks_2011}.

\item \subfind\ (Springel) is a configuration space finder \citep{subfind_2001}.

\item \voboz\ (Neyrinck) is a Voronoi tessellation based finder \citep{neyrinck_voboz:_2005}.
\end{itemize}

\section{The Data} \label{sec:data}

\subsection{Simulation Data}
The first data set used in this paper forms part of the Aquarius
project \citep{springel_aquarius_2008}. It consists of multiple dark
matter only re-simulations of a Milky Way-like halo at a variety of
resolutions performed using \gadget\ \citep[based on
\textsc{gadget2},][]{springel_cosmological_2005}.  We have used in the
main the Aquarius-A to E halo dataset at $z=0$ for this project.
This provides 5 levels of resolution, varying in complexity for which further
details are available in \citet{Onions_2012}.

The underlying cosmology for the Aquarius simulations is the same as
that used for the Millennium simulation
\citep{springel_millenium_2005} i.e. 
$\Omega_M = 0.25,\: \Omega_\Lambda = 0.75,\: \sigma_8 = 0.9,\: n_s = 1,\: h = 0.73$. 
These parameters are close to the latest WMAP data \citep{wmap_2011}
($\Omega_M = 0.2669,\: \Omega_\Lambda = 0.734,\: \sigma_8 = 0.801,\: 
n_s = 0.963,\:  h = 0.71$)
although $\sigma_8$ is a little high.  All the simulations were
started at an initial redshift of 127. Precise details on the set-up
and performance of these models can be found in
\citet{springel_aquarius_2008}.

The second data set was from the \ghalo\ simulation data
\citep{Stadel_2009}.  \ghalo\ uses a slightly different cosmology to
Aquarius, $\Omega_M = 0.237,\: \Omega_\Lambda=0.763,\:
\sigma_8=0.742,\: n_s = 0.951,\: h=0.735$ which again are reasonably
close to WMAP latest results.  It also uses a different gravity
solver, \textsc{pkdgrav2}\citep{pkdgrav}, to run the simulation
therefore allowing comparison which is independent of gravity solver
and to some extent the exact cosmology.

The details of both simulations are summarised in \Tbl{tbl:simdata}.

\begin{table}
\centering
  \caption{Summary of the key numbers in the Aquarius and
    \ghalo\ simulations used in this 
    study. $N_{high}$ is the number of particles with the
    highest resolution (lowest individual mass).  $N_{250}$ is
    the number of high resolution particles found within a sphere of
    radius 250 kpc/h from the fiducial centre at each resolution ({\it
    i.e.} those of primary interest for this study).  }
  \label{tbl:simdata}
  \begin{tabular}{l r r r }
    \hline
    Simulation & $N_{high}$ & $N_{250}$ \\
    \hline
    Aq-A-5 &  2,316,893    & 712,232      \\
    Aq-A-4 &  18,535,97    & 5,715,467    \\
    Aq-A-3 & 148,285,000   & 45,150,166   \\
    Aq-A-2 & 531,570,000   & 162,527,280  \\
    Aq-A-1 & 4,252,607,000 & 1,306,256,871\\
    Aq-B-4 & 18,949,101    & 4,771,239    \\
    Aq-C-4 & 26,679,146    & 6,423,136    \\
    Aq-D-4 & 20,455,156    & 8,327,811    \\
    Aq-E-4 & 17,159,996    & 5,819,864    \\
    \hline
    GH-4 & 11,254,149      & 1,723,372    \\
    GH-3 & 141,232,695     & 47,005,813   \\
    \hline
  \end{tabular}
\end{table}

\subsection{Post-processing pipeline}\label{ssec:pipeline}
The participants were asked to run their subhalo finders on the
supplied data and to return a catalogue listing the substructures they
found.  Specifically they were asked to return a list of uniquely
identified substructures together with a list of all particles
associated with each subhalo. The broad statistics of the haloes found
are summarised in \Tbl{tbl:ResultsCent}.

To enable a direct comparison, all the data returned was subject to a
common post-processing pipeline detailed in \citet{Onions_2012}.  For
this project we added a common unbinding procedure based on the
algorithm from the \ahf\ finder which is based on spherical unbinding
from the centre. We requested data to be returned both with and
without unbinding to allow a comparison of that procedure to feature
in this study.  Unbinding is the process where the collection of
gathered particles is examined to discard those which are not
gravitationally bound to the structure.  This common unbinding allowed
us to remove some of the sources of scatter introduced by the finders
using slightly different algorithms for removing unbound particles and
to find what difference this made to the results.

\begin{table*}
  \caption{The number of \subhalos\ containing 300 or more particles and
    centres within a sphere of radius 250kpc/h from the fiducial
    centre found by each finder after standardised post-processing
    (see \Sec{ssec:pipeline}).}
  \label{tbl:ResultsCent}
  \begin{tabular}{l *{11}{r}}
 \hline
    Name & \adaptahop&\ahf   &\ghp  &\hbt   &
    \htd&\hsd&\mendieta&\rockstar &\stf &\subfind&\voboz  \\
    \hline 
    Aq-A-5 & 24      & 23    & 23   &  23   & 18  &23  & 17      & 25       & 22  & 23     & 21 \\ 
    Aq-A-4 & 222     & 189   & 170  &169    & 174 &176 & 123     & 182      & 155 & 154    & 163 \\ 
    Aq-A-3 & -       & 1259  & 1202 &1217   &-    &-   & 787     & 1252     &1124 & 1117   & 1141 \\ 
    Aq-A-2 & -       & 4230  & -    &4036   & -   &-   & -       & 4161     & -   & 3661   & - \\ 
    Aq-A-1 & -       & 30694 & -    & -     &-    & -  & -       & 25009    & -   & 26155  & - \\ 
    Aq-B-4 & -       & 197   & -    & 191   &-    & -  & -       & 202      & -   & 188    & - \\ 
    Aq-C-4 & -       & 152   & -    & 146   &-    & -  & -       & 158      & -   & 137    & - \\ 
    Aq-D-4 & -       & 217   & -    & 216   &-    & -  & -       & 230      & -   & 196    & - \\ 
    Aq-E-4 & -       & 218   & -    & 219   &-    & -  & -       & 221      & -   & 205    & - \\ 
    GH-4  & -       & 58    & 58    & -     &-    & -  & -       & 60       & 54  & 54    & - \\ 
    GH-3  & -       & 1172  & 1148  & -     &-    & -  & -       & 1148     & 1033& 1090  & - \\ 

    \hline 
  \end{tabular}
\end{table*}

Both the halo finder catalogues (alongside the particle ID lists) and
our post-processing software are available from the authors
on request.

\begin{figure}
 \centering
 \includegraphics[width=1\linewidth]{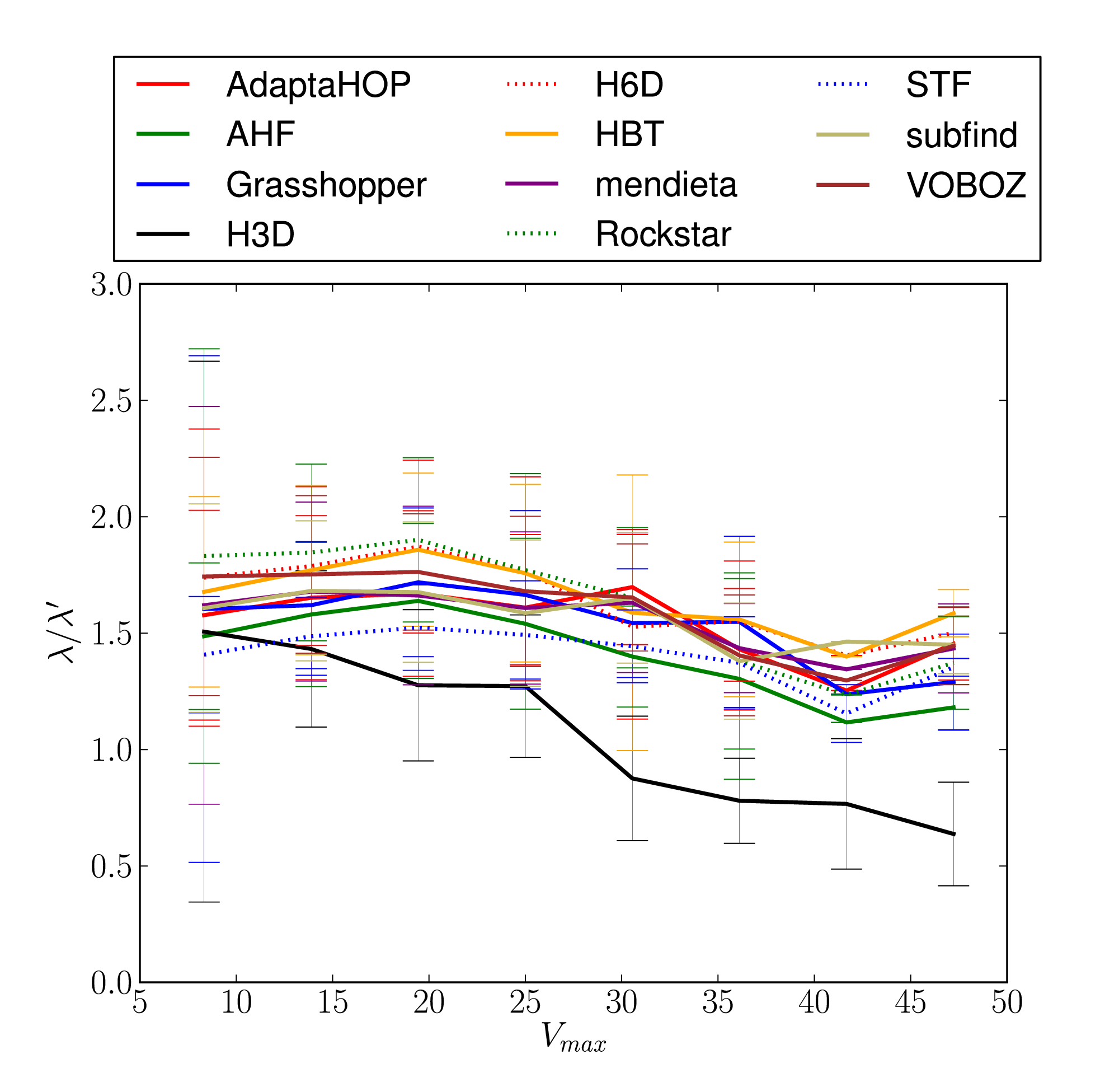}
\caption{A comparison of the Peebles and Bullock spin parameters against
\vmax\ based on all finders using a common unbinding procedure from
subhaloes with more than 300 particles. The mean
value of $\lambda/\lambda'$ is shown together with one standard deviation
error bars.  It shows there is a correlation between the two but not a
one-to-one correspondence, with some scatter present. The scatter
at low \vmax\ where haloes have very few particles is particularly pronounced.}

\label{fig:spinPvB} 
\end{figure}

\section{Results} \label{sec:Results}

The results used were restricted to subhaloes with more than 300
particles, as these produce a relatively stable value for
spin. Values below this limit tend not to converge across resolutions
\citep{bett_2007}.

\subsection{Spin parameter}\label{sec:spin}

In general there is a proportional relationship between the Peebles and
Bullock spin parameters recovered by all the finders for the same
subhaloes, although there is some scatter as shown in
\Fig{fig:spinPvB}.  We do not dwell on the differences between the two
definitions as that has already been studied elsewhere
\citep{hetznecker_2006}.  As both definitions of spin exist in the
literature we consider both metrics when comparing how the spin is
recovered across finders, placing particular emphasis on their
application to subhaloes.

The majority of
field haloes are found to cluster around a value of $\lambda_0=0.044$
for the Peebles spin parameter \citep{bett_2007} and $\lambda_0'=0.035$
for the Bullock parameter \citep{bullock_2001} with a spread of values
matched by a free parameter to give the width of the distribution.

\begin{figure}
 \centering
 \includegraphics[width=0.49\linewidth]{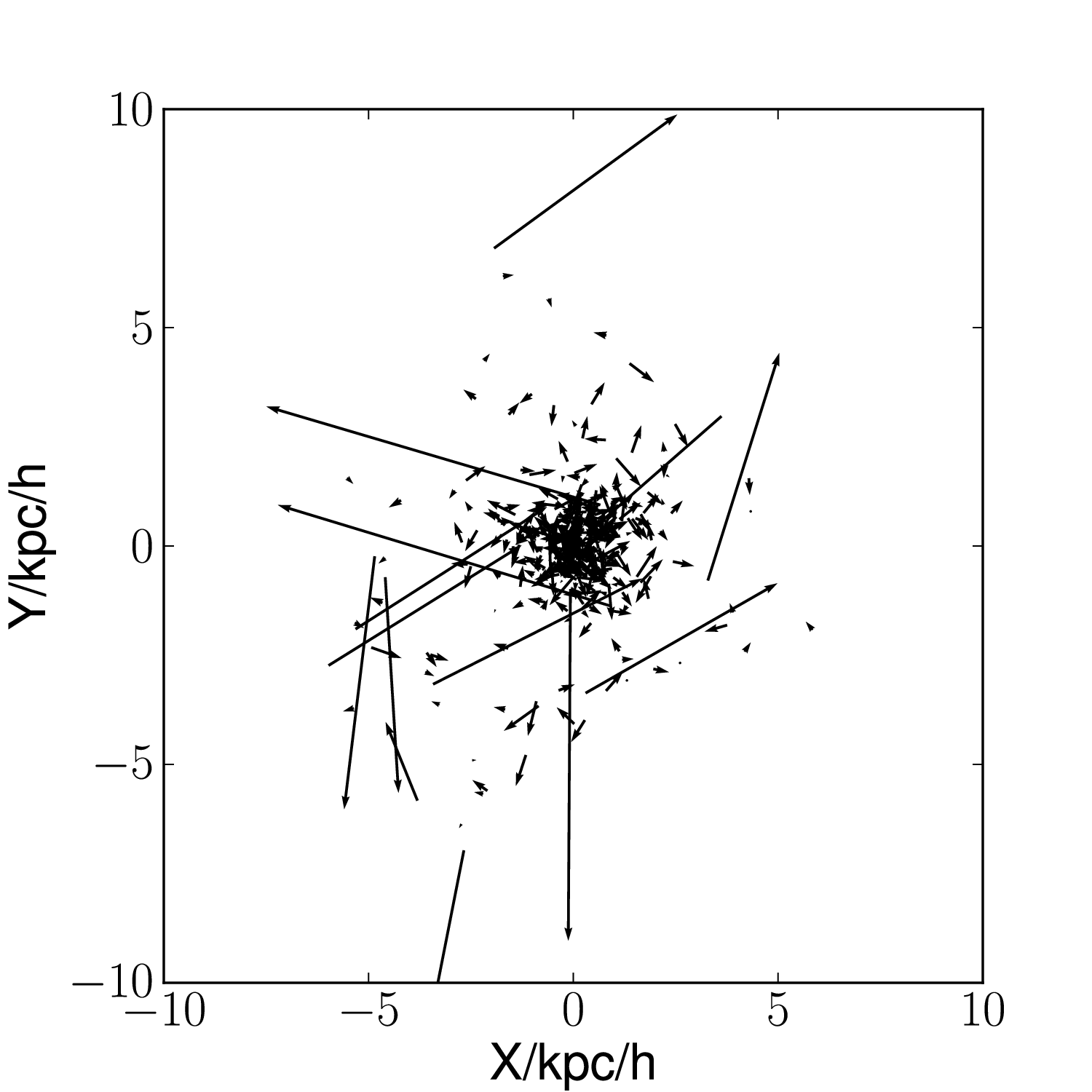} 
\includegraphics[width=0.49\linewidth]{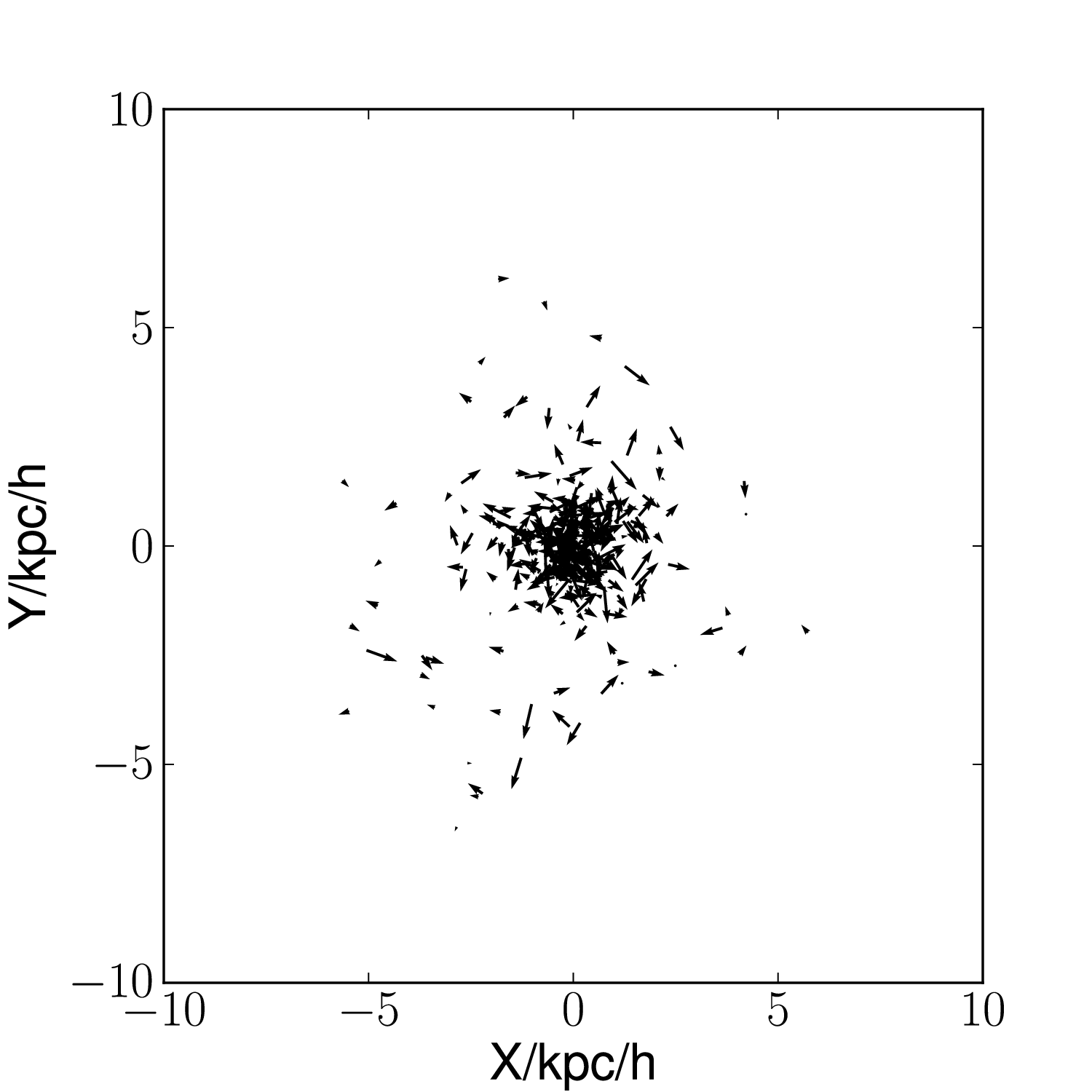}
\caption{An example of the influence of unbinding. Left panel:
  particles in the object prior to unbinding. Right panel: particles
  in the object after unbinding the been performed. The vectors
  indicate the direction and velocity relative to the bulk velocity of
  the individual particles making up this example subhalo.  The
  contribution from the background particles has only a minor
  influence on the mass and \vmax\ of the subhalo, but a large effect
  on the spin parameter.  }
\label{fig:unbindexample}
\end{figure}

\subsubsection{Spin for subhaloes with no unbinding performed}
If unbinding has not been correctly implemented the high speed
background particles can distort the spin parameter
enormously.

To emphasise the type of structures that are found, an example of a
subhalo without (left panel) and with (right panel) unbinding is shown
in \Fig{fig:unbindexample}. This is displayed as a vector plot of all
the component particles position and velocities that make up the
subhalo with the velocity vectors scaled in the same way in both
panels. The bulk velocity of the subhalo has been removed and all
positions and velocities are relative to the rest frame of the
subhalo. Evident in the left panel of \Fig{fig:unbindexample} without
unbinding are stray particles that are part of the background
halo. Despite their small number these particles have both a large
lever arm and large velocity relative to the halo, and significantly
alter the derived value of the spin parameter due to their large
angular momenta.

\begin{figure*}
 \centering
\begin{minipage}[t]{0.48\linewidth}
 \includegraphics[width=1\linewidth]{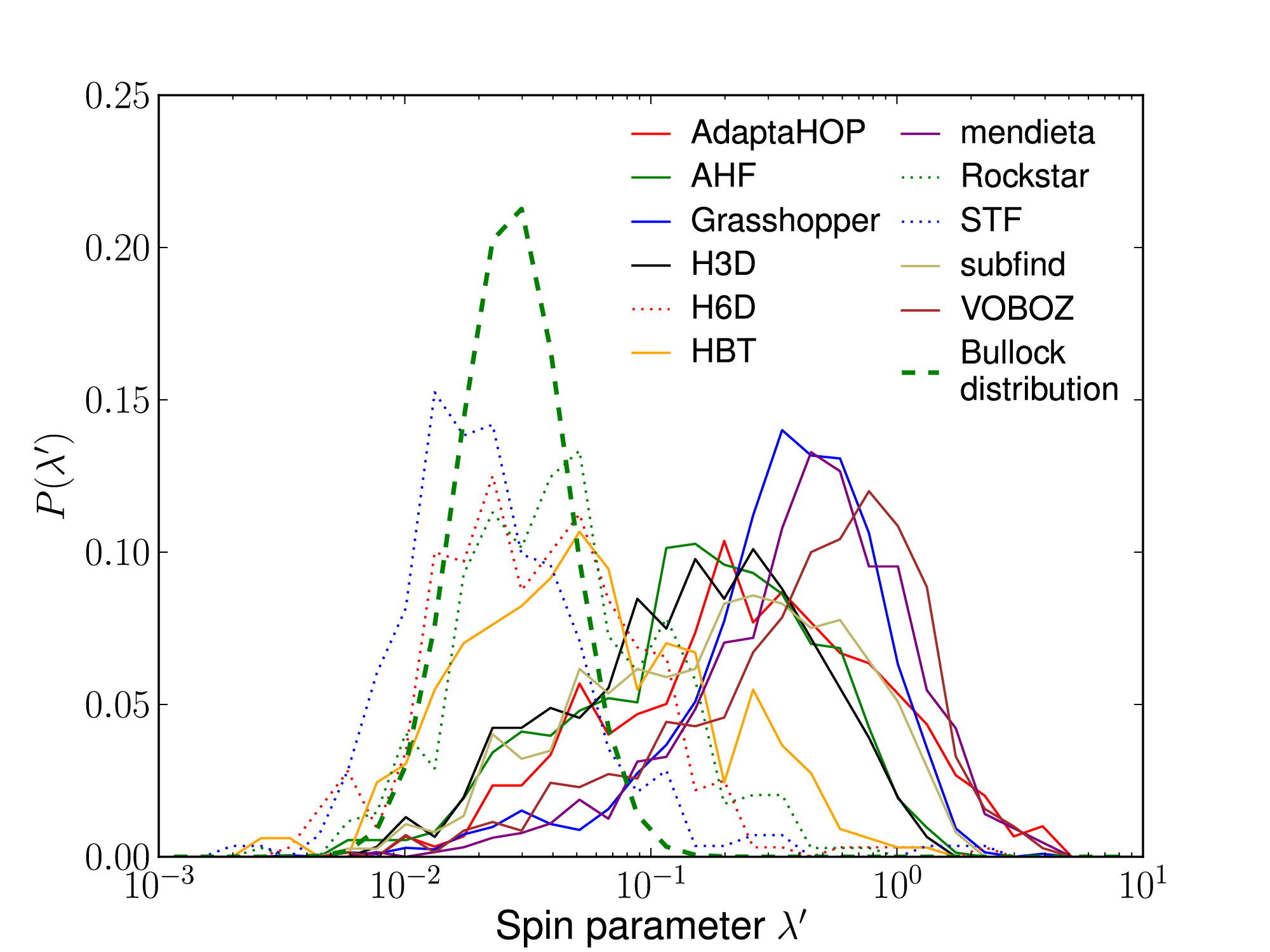}
\caption{General profile of the Bullock spin parameter of all
subhaloes found with more than 300 particles without unbinding
performed, binned into 35 log bins. The results are 
normalised to give equal area under the visible curve.
The dashed line is the field halo
fit from \citet{bullock_2001}.  The results show a large scatter about
a peak which is far distant from the fiducial fit for haloes. Dotted lines
indicate finders with a phase space component of their algorithm,
whereas solid lines indicate finders without a phase space component.
}
\label{fig:bullockraw}
\end{minipage}\hskip \columnsep
\begin{minipage}[t]{0.47\linewidth}
 \centering
 \includegraphics[width=1\linewidth]{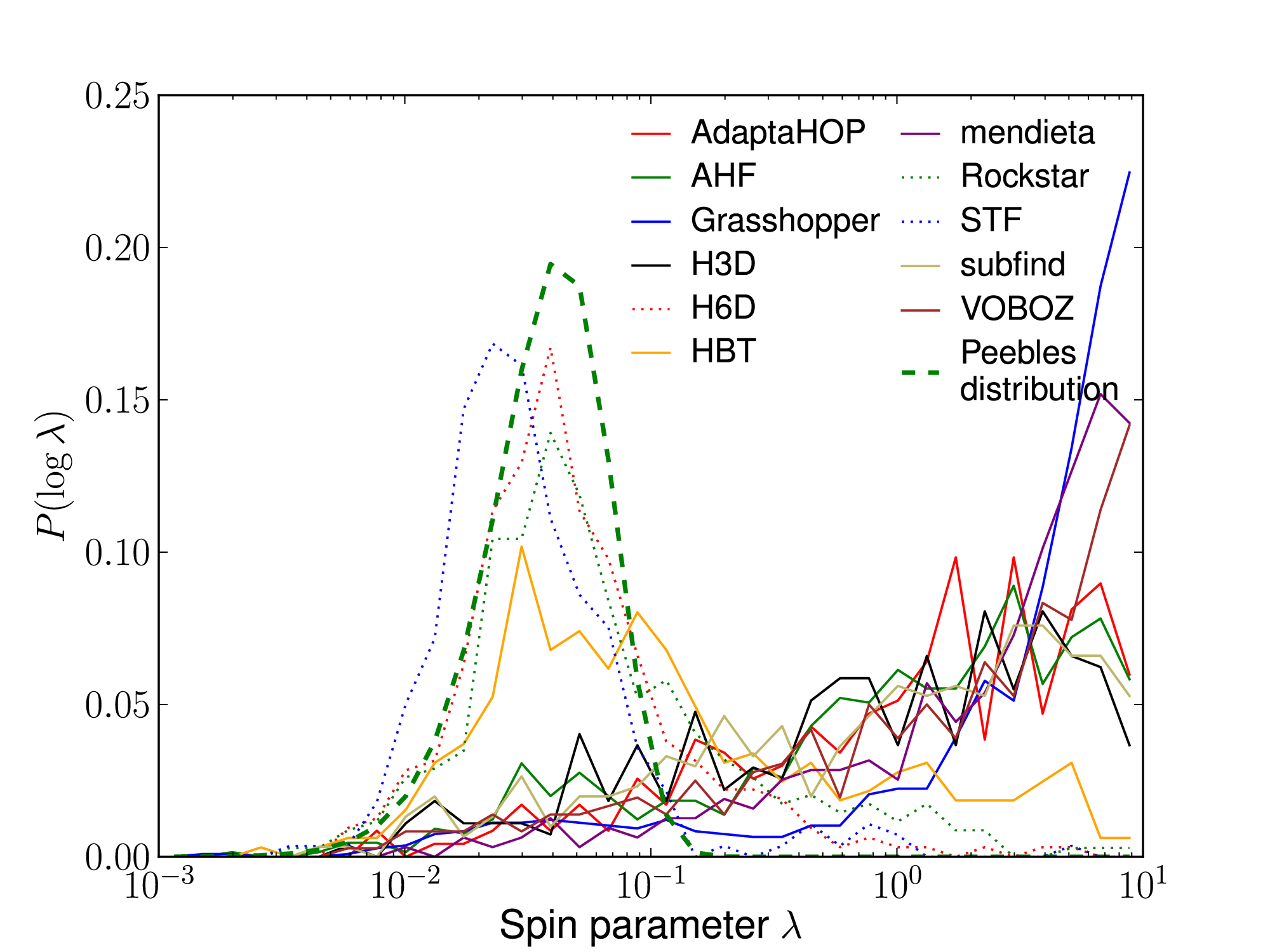}
\caption{The same plot as \Fig{fig:bullockraw} but using the Peebles
  spin parameter and fitting function from \citet{bett_2007}.  }
\label{fig:peeblesraw}
\end{minipage}

\begin{minipage}[t]{0.47\linewidth}
 \centering
 \includegraphics[width=1\linewidth]{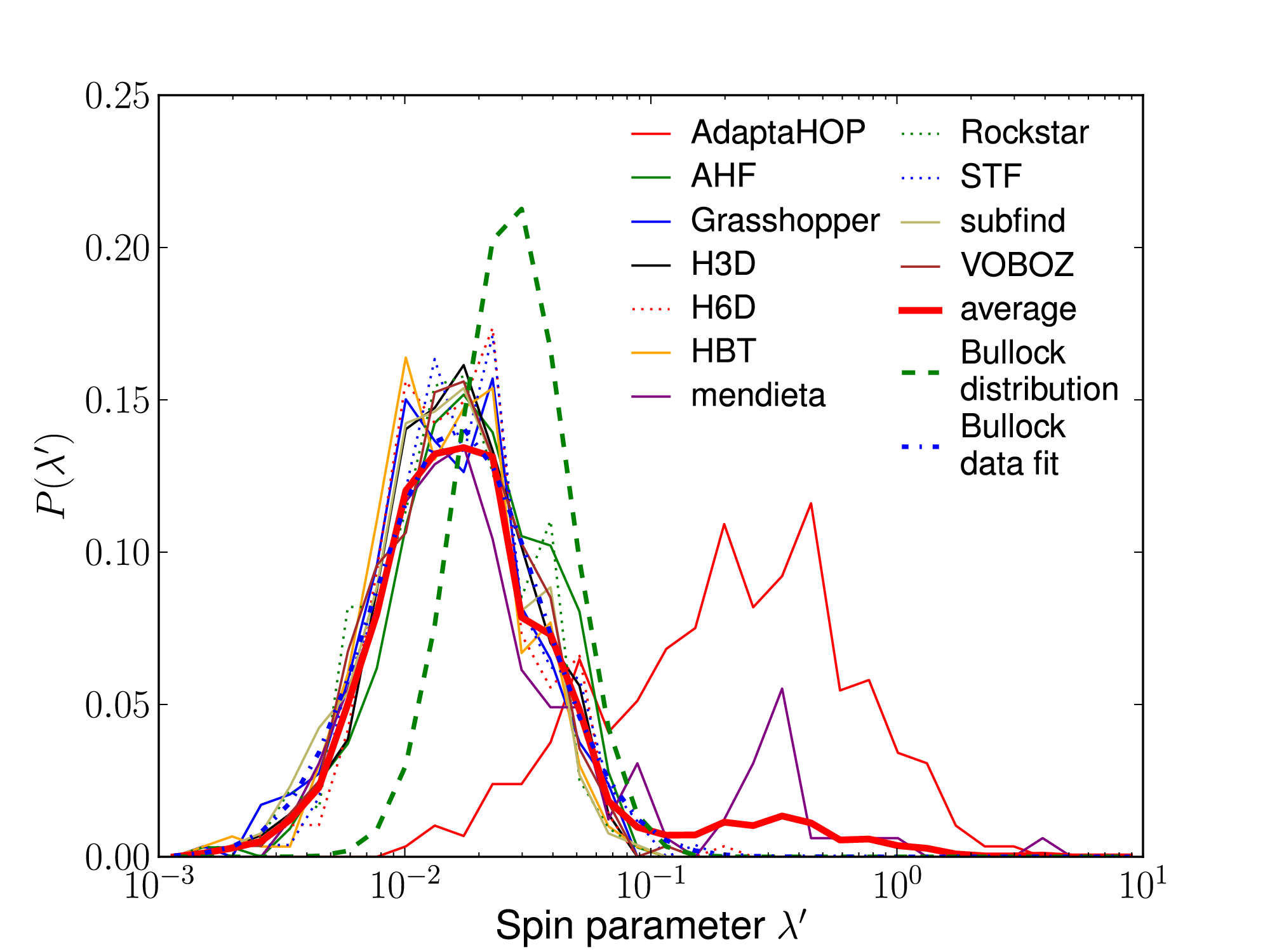}
\caption{The same plot as \Fig{fig:bullockraw} but with the finders
own unbinding processing applied to the data. This groups the spin
parameters somewhat more tightly, and shows that spin is a good
indicator of how well the unbinding procedure is removing
spurious background particles. The \adaptahop\ finder doesn't perform
an unbinding step, and this plot also shows up a flaw in \mendieta's
unbinding procedure.  The dashed line is the Bullock field halo fit
curve from \citet{bullock_2001}. The Bullock data fit is the best
fit to the average using the Bullock fitting formula. }
\label{fig:bullockownunbound}
\end{minipage}\hskip \columnsep
\begin{minipage}[t]{0.47\linewidth}
 \centering
 \includegraphics[width=1\linewidth]{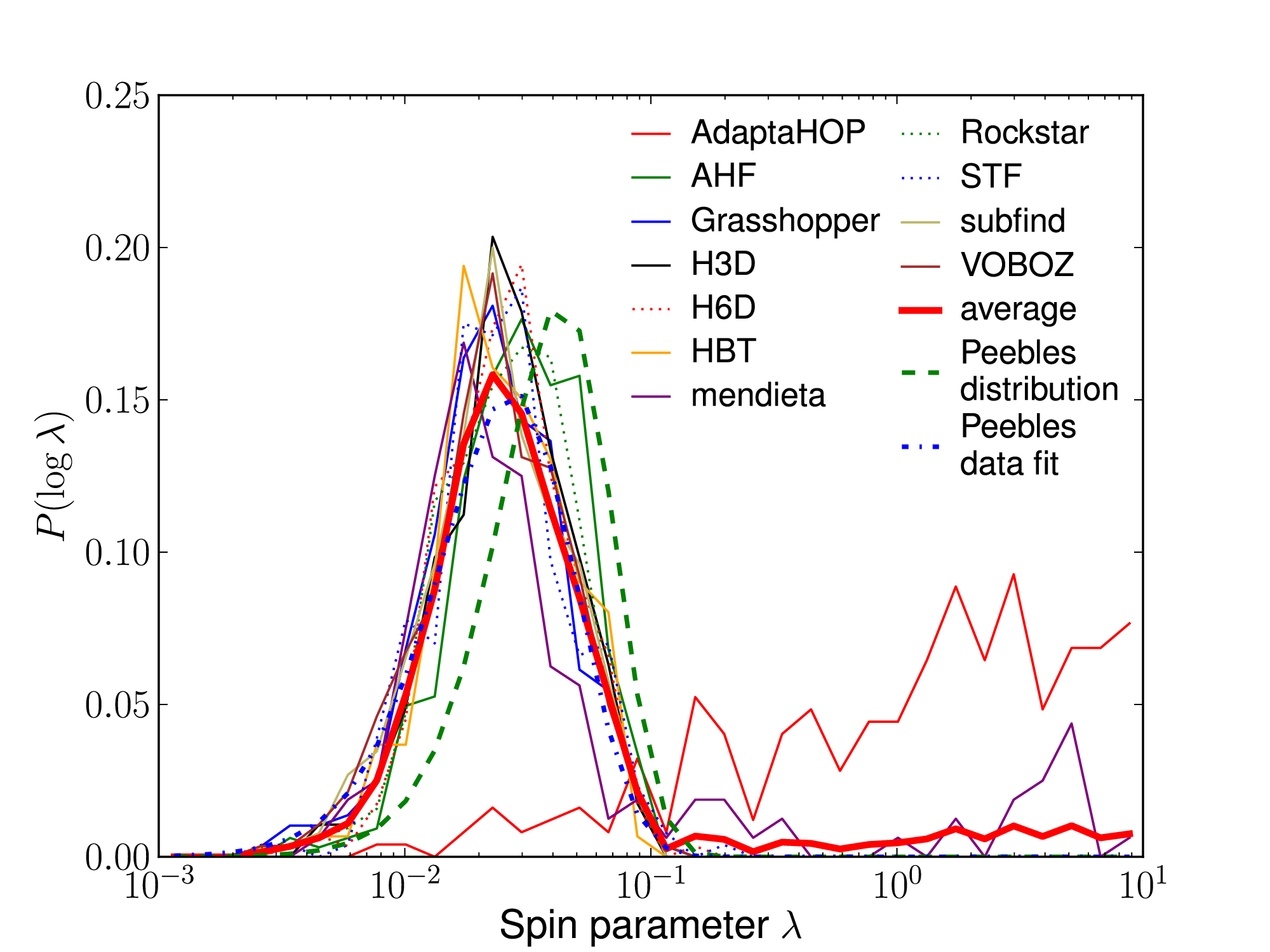}
\caption{The same plot as \Fig{fig:bullockownunbound} except that this
  time the dashed line is the Peebles field halo fit from
\citet{bett_2007}. The Peebles best data fit is the best fit to the average
of the Bett formula. }
\label{fig:peeblesownunbound}
\end{minipage}
\end{figure*}

Comparing the two forms of the spin parameter in
\Fig{fig:bullockraw} and \Fig{fig:peeblesraw} we show how the spin
parameter is quite chaotic, not matching a smooth Gaussian like profile as 
might be expected, and is clearly a long way removed from the
idealised curve others have found for the distribution of spin.
A significant number of the haloes
have spin parameter values above 1, which is unphysical as these
objects would be ripped apart by this level of rotation and so clearly
cannot be equilibrium systems. This result is perhaps not surprising
given the contribution from unbound background particles moving with
velocities far from the mean of the object being considered but
clearly shows how poor unbinding methods are relatively easy to detect
by looking at the spin parameter distribution.  The Peebles spin
parameter is more affected by the lack of unbinding than the
equivalent Bullock parameter as it takes into account the kinetic
energy of all the particles. Some more objective numbers for this and
subsequent comparisons are given in \Tbl{tbl:bindparams}.

The best fit values shown by the bold dashed lines are vastly
different from the fiducial values given in \Sec{sec:method}. It is
however significant that the finders \hsd, \rockstar\ and \stf\ (shown
by dotted lines) which all have a phase space based component in their
particle collection algorithm already show a much better fit to the
fiducial value than the non phase-space finders.  It should be
noted that when \ghp\ is run without unbinding, it finds a 
large number of subhaloes which would
normally be discarded by the unbinding procedure that is
integral to the final part of the \ghp\ algorithm.

\begin{figure*}
\begin{minipage}[t]{0.47\linewidth}
 \centering
 \includegraphics[width=1\linewidth]{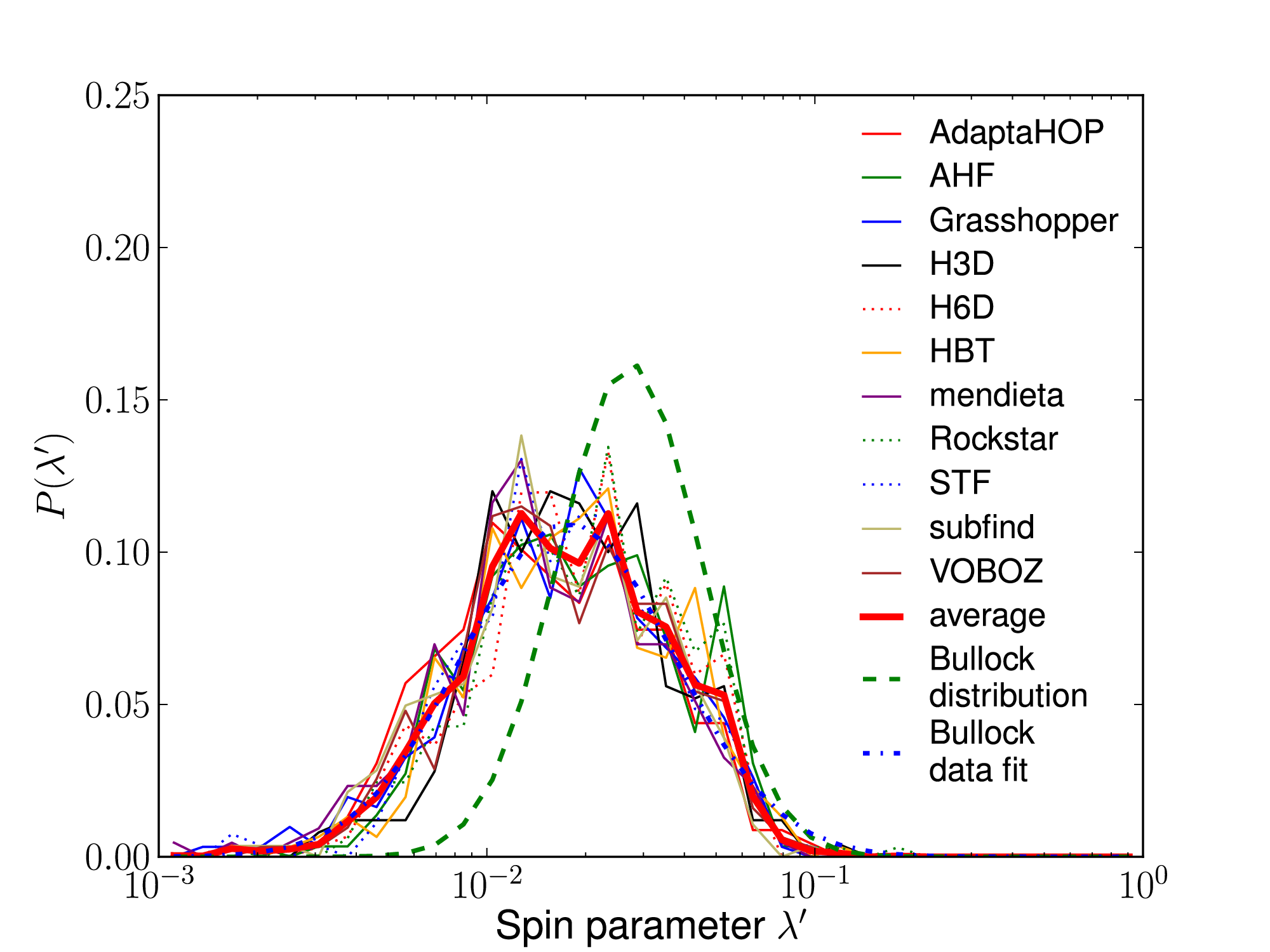}
\caption{The same plot as \Fig{fig:bullockraw} but with a common
unbinding processing applied to the data. This groups the spin
parameters much more tightly, and shows that spin is a good predictor
of how well the unbinding procedure performing at removing spurious
background particles.  The dashed line is the Bullock best fit field halo curve
from \citet{bullock_2001}.  }
\label{fig:bullockunbound}
\end{minipage}\hskip \columnsep
\begin{minipage}[t]{0.47\linewidth}
 \centering
 \includegraphics[width=1\linewidth]{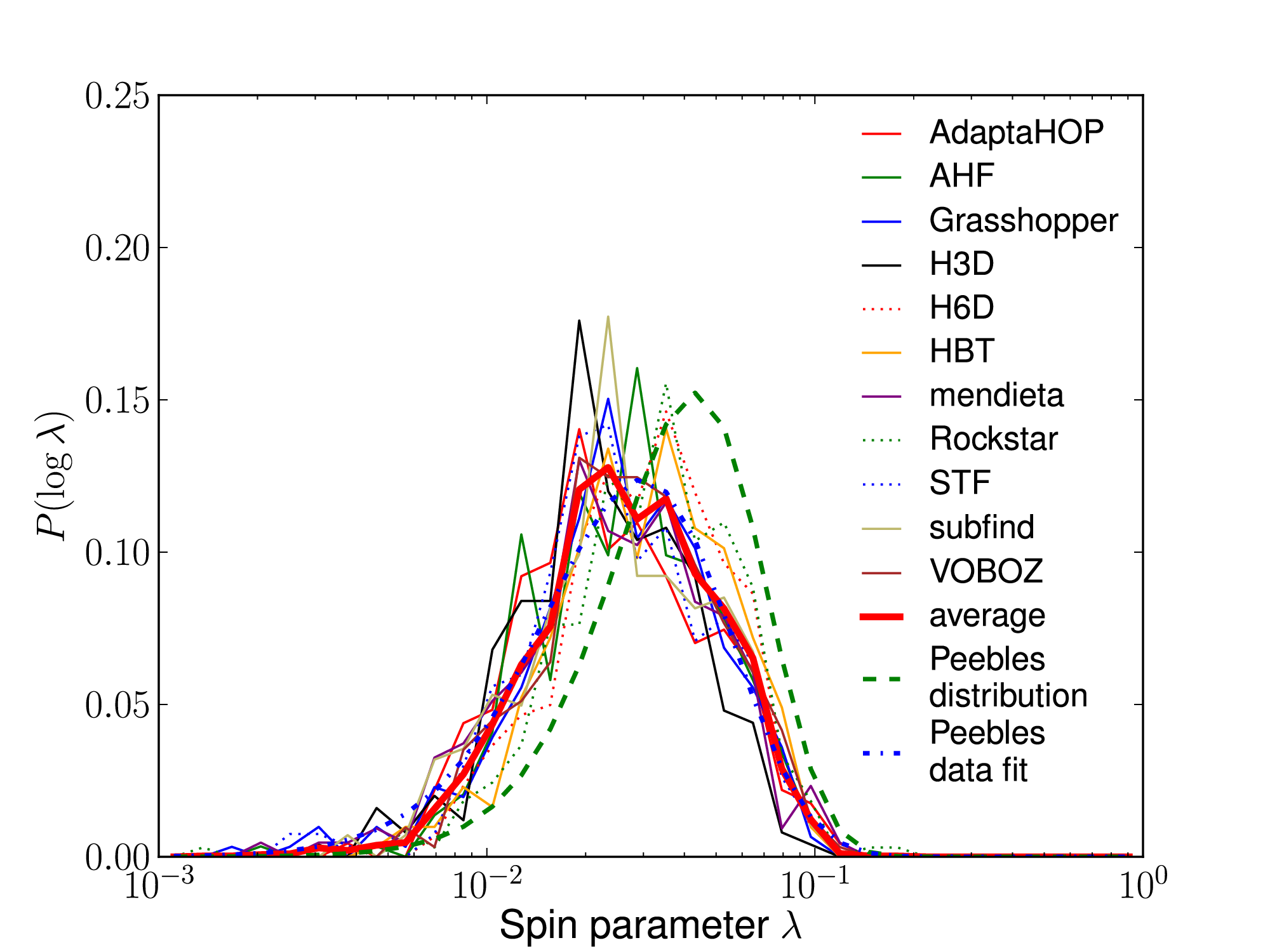}
\caption{The same plot as \Fig{fig:peeblesraw} but with a common
unbinding processing applied to the data.  The dashed line is the
Peebles best fit curve from \citet{bett_2007}.  }
\label{fig:peeblesunbound}
\end{minipage}
\end{figure*}
\subsubsection{Spin for subhaloes with finders own unbinding performed}

Including each finder's own unbinding procedure improves the spin
parameter measure considerably, as shown in
\Fig{fig:bullockownunbound} and \Fig{fig:peeblesownunbound}.  Note
that as \adaptahop\ doesn't do any unbinding in its post-processing
steps it is a clear outlier on this plot.  The \mendieta\ finder shows
a double peak, which is indicative of some of the unbinding failing,
an issue that the authors of the finder are currently working on.

When fitting the best fit curves to this data obtained for the spin
parameter of subhaloes, the peak of the Bullock fitting curve given in
\Eqn{eqn:bullock} is less than the field halo value by about 20 percent,
offsetting the mean towards smaller values of the spin parameter.  For
the Peebles spin parameter the best fit is again offset by about 36
percent from the field halo value, again towards a smaller value of the
spin parameter.

\subsubsection{Spin for subhaloes with a common unbinding performed}

Once a common unbinding is done, the curves move significantly closer
to the idealised curve, although there is still some separation.  The
plots of \Fig{fig:bullockunbound} and \Fig{fig:peeblesunbound} compare
the spin parameter distribution of the different finders using a
common unbinding process. It shows the match between the best fit
curve quoted in \cite{bullock_2001} and \citet{bett_2007} and the
haloes found by the finders taking part in the comparison.  The values
are now offset by 10 percent for the Bullock fit, and 30 percent for
the Peebles fit. This results in the closest fit to the data, although
the subhalo spin again extends to slightly lower values for both
parameters, and follows the best fit line at larger values.  These
results also have a similar trend for the Aquarius B-E haloes and
the \ghalo\ data sets. These inclusions show that the results are
not influenced greatly by the simulation, simulation engine or small
changes in the cosmology used.

\subsubsection{Spin at higher resolutions}

\begin{figure}
 \centering
 \includegraphics[width=1\linewidth]{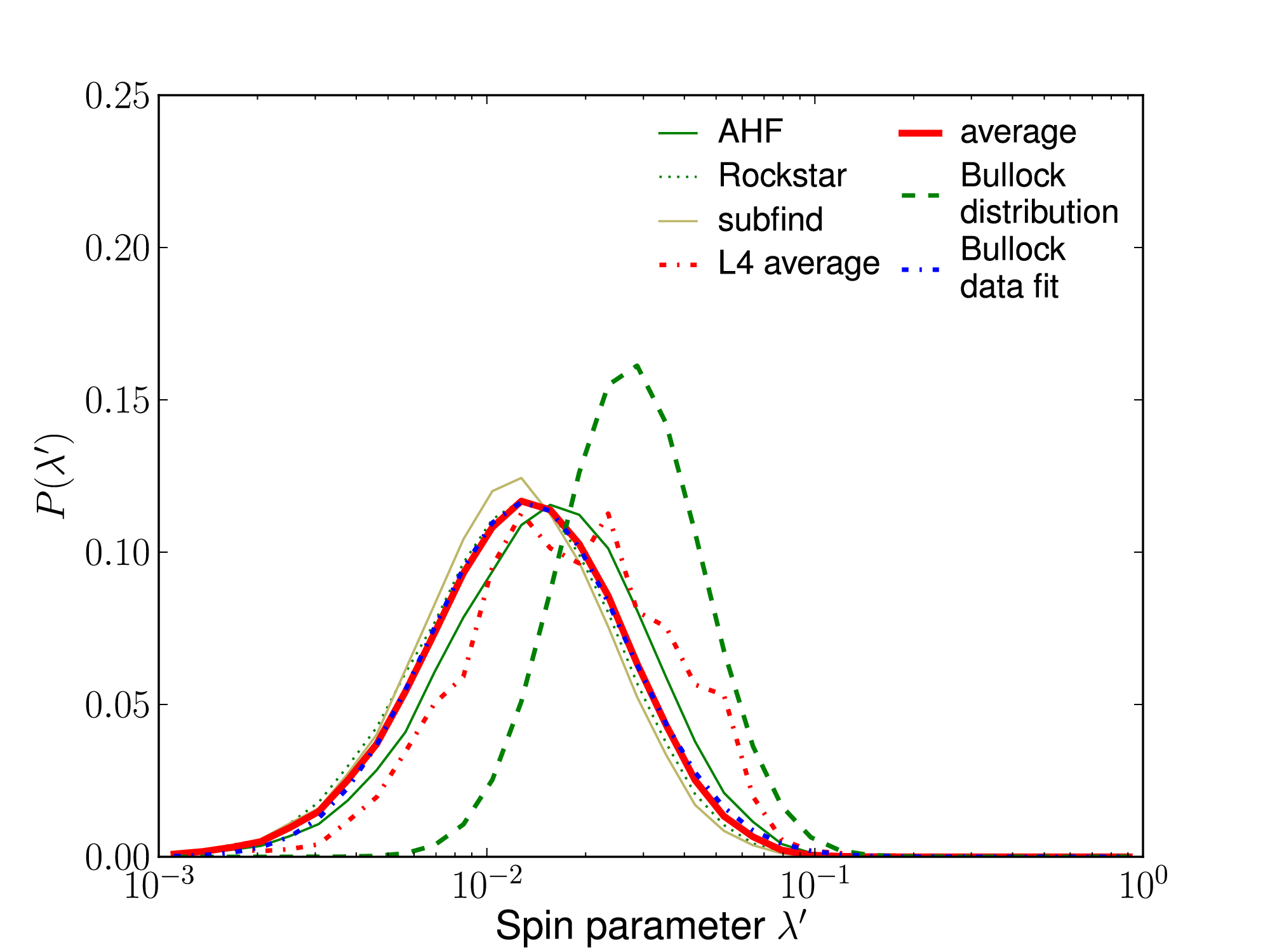}
\caption{The same plot as \Fig{fig:bullockownunbound} but using
the level 1 data which has much higher resolution. The lower spin
haloes are more obvious in this plot, as is the difference between
finders. The level 4 average is included for comparison.
}
\label{fig:bullockownl1}
\end{figure}

Going to higher resolutions afforded by the level 1 data as shown in
\Fig{fig:bullockownl1}, the trend to a lower spin distribution peak
continues, although only three of the finders were able to manage such
a computationally intensive task.

There is a more pronounced tendency to depart from the field
halo fit line at low spin part of the distribution, with the peak and bulk of
the distribution moving towards lower spin parameter values. 
The finders also show more scatter with each of them identifying 
the peak of the distribution in slightly different places.
The agreement particularly at the low end of the spin distribution is good
but with slightly lesser agreement at the high end. 

Although \ahf\ appears
to find slightly more higher spin haloes, this is a result of the
spherical unbinding algorithm it uses, which tends to also increase the
spin distribution of the other finders slightly when used as the common
unbinding procedure.

The dashed line representing the level 4 data is included to allow a
direct comparison between the level 4 and level 1 average fits. It
shows the continued movement of the distribution towards lower spin
values with higher resolution and an increase in data.

\subsubsection{Spin distribution summary}

The best fit curve figures for all these plots are summarised in
\Tbl{tbl:bindparams}. Even after cleaning the catalogues significantly
by utilising a common unbinding procedure for all finders there
remains a definite trend for substructure spin to be less than that
found for field haloes. We investigate the reason for this in the next
sections.

\begin{table}
  \caption{Summary of the best fit parameters for the graphs
shown. Shown are the values for $\lambda_0$ and the other free
parameter ($\alpha$ or $\sigma$) used in the best fit, and their
difference from the published field halo fit value.  The subscripts F, N,
O and C are for field haloes, no unbinding, own unbinding and common
unbinding respectively. The $\Delta$ values are the difference from
the field halo values, and the change is the percentage difference.
All results are for level 4 data except the last which is level 1}
  \label{tbl:bindparams}
  \begin{tabular}{l l l l l l l}
 \hline
Plot & $\lambda_0$ & $\Delta\lambda_0$ & change & $\sigma / \alpha $ & $\Delta\sigma / \alpha$ & change \\
 \hline
Bullock$_F$ & 0.035 &        &         & 0.5   &       & \\
Peebles$_F$ & 0.044 &        &         & 2.509 &       & \\
Bullock$_N$ & 1.646 & 1.611  & +4600\%  & 1.36  & 0.86  & 172\% \\
Peebles$_N$ & 12.6  & 12.573 & +29000\% & 41    & 39.2  & 1560\% \\
Bullock$_O$ & 0.028 & -0.007 & -20\%   & 0.727 & 0.227 & 45.5\% \\
Peebles$_O$ & 0.028 & -0.016 & -36\%   & 3.643 & 1.134 & 45.2\% \\
Bullock$_C$ & 0.031 & -0.004 & -10.4\% & 0.75  & 0.25  & 50.0\% \\
Peebles$_C$ & 0.03  & -0.013 & -30\%   &3.96   & 1.448 & 57.7\% \\
Bullock-L1$_O$ & 0.022 & -0.013 & -38\%   & 0.693 & 0.193 & 38.6\% \\
 \hline
  \end{tabular}
\end{table}

\subsection{Host halo radial comparison}\label{ssec:radcomp}

\begin{figure}
 \centering
 \includegraphics[width=1\linewidth]{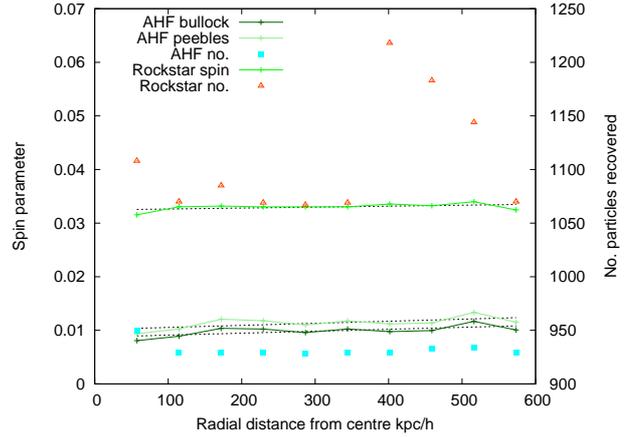}
\caption{The spin parameter as recovered by \ahf\ (Peebles and Bullock)
and \rockstar\ (Peebles) of an
outer  subhalo repositioned progressively closer to the centre. The
finders own spin calculations were used in this case rather than the
full pipeline.  The spin is seen to be approximately unchanging across
the radius.}
\label{fig:droppedhalo} 
\end{figure}

Next we consider whether the location of a subhalo within a host halo
has any effect on the recovered spin parameter. First we demonstrate
in \Fig{fig:droppedhalo} that any effect is not an artefact of the
finding process.  Substructures closer into the centre of the host halo
are more difficult to detect particularly by some finders, and therefore
subject to a loss of constituent particles that could be attached to
the subhalo as shown in \citet{muldrew_accuracy_2011}.  To test this
supposition we took a subhalo found in the outskirts of the Aquarius-A
main halo, and repositioning it at points closer to the location of the
centre of the halo. Then two of the finders (\ahf\ and \rockstar) were
rerun on the new data and the spin value calculated anew. The results
shown in \Fig{fig:droppedhalo} indicate that there is little change in
the value of the spin parameter with radius despite some variation in
the recovered number of particles.

\begin{figure}
 \centering
 \includegraphics[width=1\linewidth]{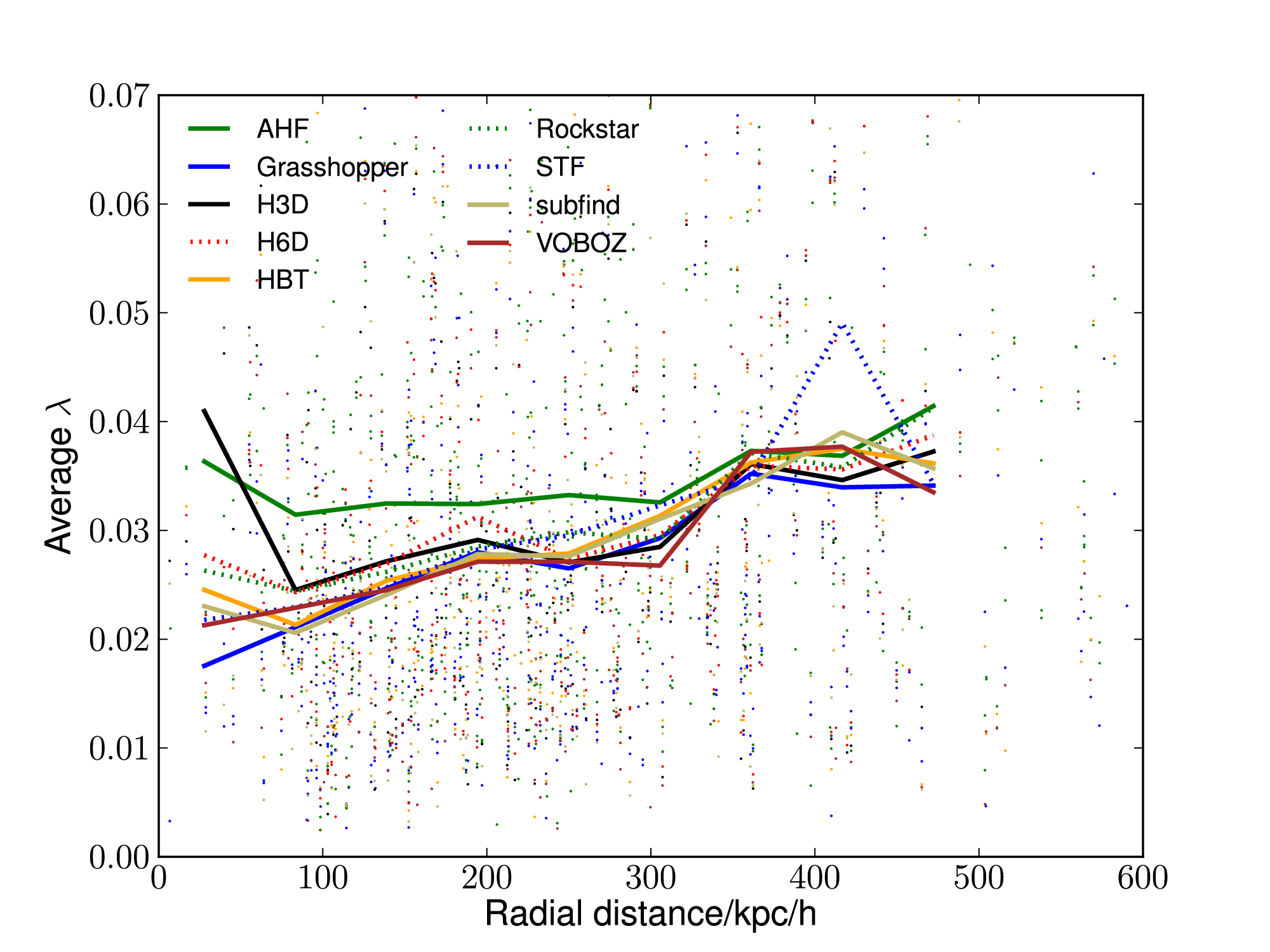}
\caption{Comparison of mean spin parameter against radius from the 
centre of the host halo. Common unbinding was applied
in the pipeline in this case. There is some additional scatter
at low radial values as few haloes above 300 particles are found there.
The background points indicate the measured spin parameter for
individual subhaloes.}
\label{fig:radialspinunbinding}
\end{figure}

Next we look at whether the mean value of the measured spin parameters
changes with respect to the distance from the centre of the host
halo. \Fig{fig:radialspinunbinding} displays this radial dependence
for the indicated finders after a common unbinding step has been
applied. The background points indicate the scatter in the spin
parameter for any individual halo, as seen in the previous
section. This shows a small trend for a lower mean spin as the
subhaloes get closer to the centre of the host halo. 
This confirms the result that were found in \citet{Reed_2004} but is shown here
at higher resolution and across more finders than the earlier paper.

\begin{figure}
 \centering
 \includegraphics[width=1\linewidth]{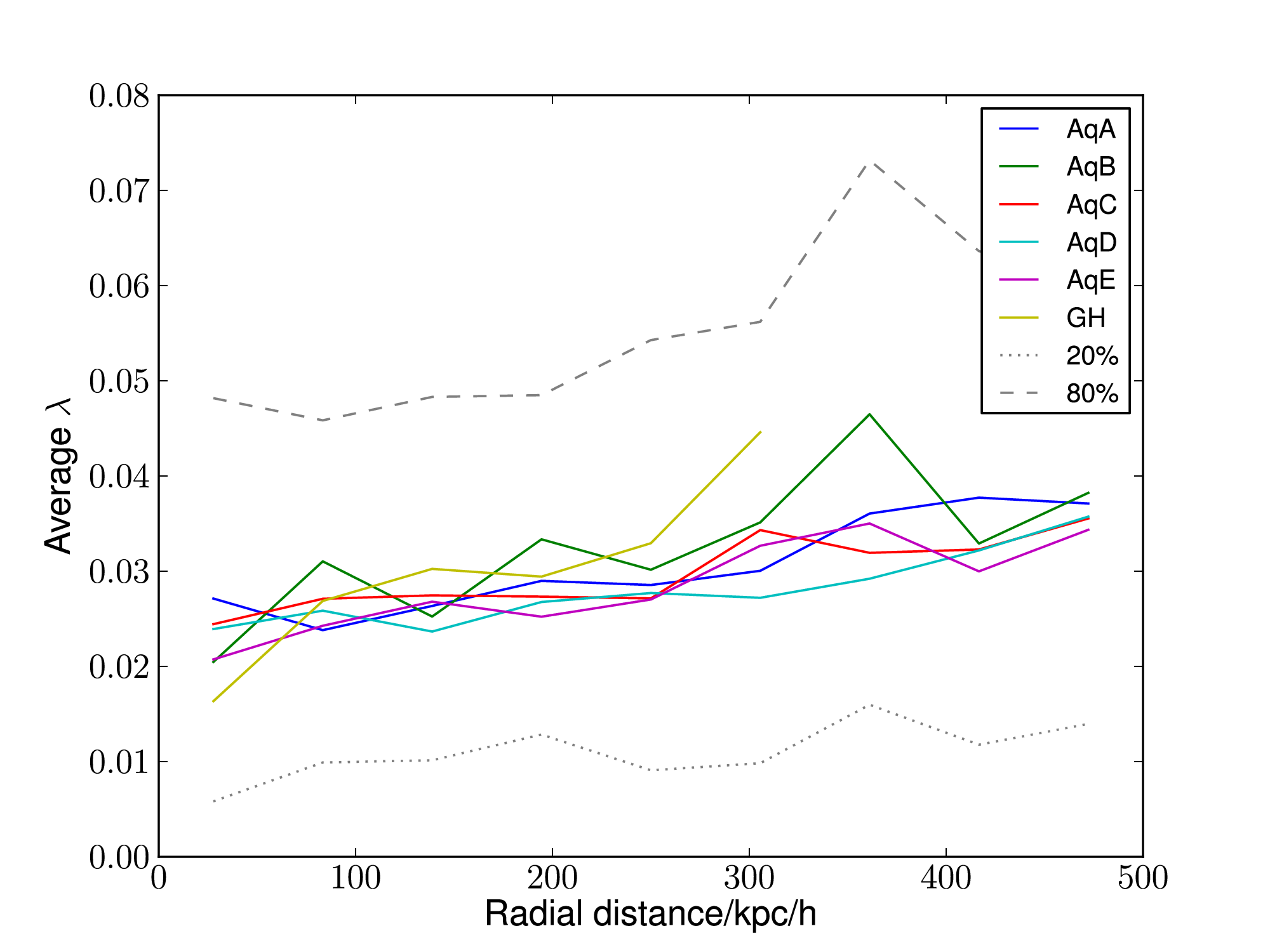}
\caption{Comparison of mean spin parameter against radius from the centre
of the host halo for several different haloes. The finders own
unbinding procedure was used in the pipeline in this case. Each line
is the average of the spin parameter binned into 10 bins across all
finders partaking (\ahf, \ghp, \rockstar, \subfind and \stf).  The
haloes used were the Aquarius-A to E and \ghalo\ all at level 4 of the
resolution. The dashed/dotted lines indicate 20 and 80 maximum percentiles 
across all data.}
\label{fig:radialspinaverage}
\end{figure}

Equivalent results are found when we compare 6 different
simulations generated by two different N-body codes and
aggregate the average of the different finders across multiple
haloes in \Fig{fig:radialspinaverage}. This effect (as noted in
\citealt{Reed_2004}) is difficult to detect observationally, as most
substructure will form galaxies before falling in so will have its spin
detectable from observations of galactic rotation curve already fixed
\citep{kauffmann_1993}. The possible exception to this are galaxies
forming at high redshift where the infalling substructure has not yet
formed stars, such as gas-rich dark galaxies \citep{cantalupo_2012},
made entirely of dark matter and gas, which may form structure after
falling into a parent halo.

\subsection{Build up of the spin parameter within a subhalo}\label{ssec:submass}
\begin{figure}
 \centering
 \includegraphics[width=1\linewidth]{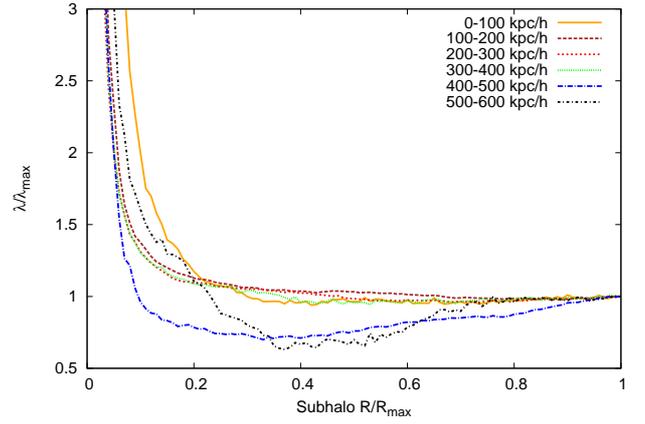}
\caption{The radial profile of the spin parameter across the 
  the subhalo.  This shows the change in the measured spin parameter as
  spin is analysed from the centre to the radius of the subhalo.  Here $R_{max}$ is the subhaloes
  maximum radius.  Each line represents a different host halo radial bin.
  Subhaloes near the centre of the host halo show monotonically rising
  spin parameter values spin, whereas further out the spin parameter
  initially drops before rising.  }
\label{fig:radialbinspin} \end{figure}

This leads to the question of what causes the drop in the measured
spin parameter with proximity to the centre of the host halo.
\Fig{fig:radialbinspin} shows the average change in the measured spin
parameter as the detected subhalo is analysed from the centre outwards to
its radius. This procedure is computed after the common processing and 
unbinding steps have been done. 
The subhaloes analysed in this way are then further 
binned into radial bins determined from the centre of the host halo.
The outermost subhaloes, which are the least disrupted, show an initial
decrease in measured spin parameter as particles are removed from their
outer edges. Subhaloes extracted from nearer the centre of the host halo
do not show this initial decrease but instead have a monotonically
rising spin parameter as material is removed.

This trend suggest that subhaloes are preferentially stripped of high
angular momentum particles which are likely to be the most weakly bound
particles, leading to a decrease in the spin parameter as they
enter the host halo. The outermost particles are usually those least
bound so are the most likely to be removed on infall.

\begin{figure}
 \centering
 \includegraphics[width=1\linewidth]{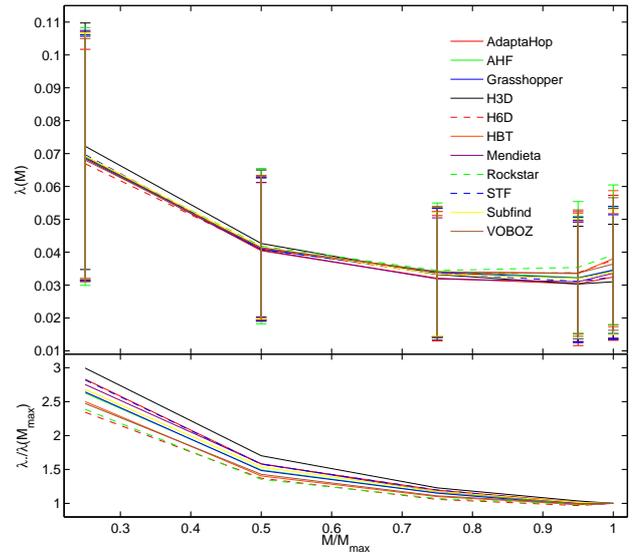}
\caption{Comparison of the normalised mean Peebles spin at different
mass shells of all subhaloes.  The cuts were taken at 0.25, 0.5, 0.75,
0.95 and the complete mass of the subhalo. A common unbinding procedure
was run on the results. There is a clear decrease in spin with increasing
contained mass, and about a 3-fold drop is evident.  The top plot shows
the value of the spin parameter, and the bottom plot the spin parameter
normalised to the value of $\lambda$ at the subhaloes $R_{max}$.  Error
bars are one standard deviation.  } 
\label{fig:massshells} 
\end{figure}

We can also examine how the spin parameter is built up as mass is
added to a subhalo. In \Fig{fig:massshells} we look at how the spin
parameter changes at various mass cuts of the subhalo,
$M(<M_{tot})$. This shows how the spin is built up across the
structure of the subhalo. For each halo we calculate the spin
parameter at 0.25, 0.5, 0.75, 0.95 of the subhalo's total mass for all
the contributing halo finders. We plot the mean and the standard
deviation at each mass cut.

As expected from \Fig{fig:radialbinspin} all finders agree
that the calculated spin increases as the fraction of the subhalo mass that
is used to calculate the spin parameter is reduced. Note that haloes
have steeply rising density profiles and so the inner $50\%$ of the
mass is contained within a much smaller fraction of the radius and
that this result is averaged over all the recovered haloes and not
split in radial bins.

\section{Summary \& Conclusions} \label{sec:Summary}
There is a good level of agreement amongst the finders on the recovery
of the distribution of the spin of subhaloes, although differences are
still evident, causing scatter in some of the comparisons. 
Undoubtedly some of the scatter is due to different types of subhalo
that are being recovered by the finders, some finders focusing on 
stream like structures and some on simple overdensities. There is
still some room for improvement of the finders as the common unbinding
test shows. Some of the possible improvements and sources of error will be
outlined in Knebe et al. (in prep).

The distribution of spin provides a very good indicator of the finders
unbinding ability and seems broadly unaffected by the cosmology and
simulation engine in use. As such, the spin distribution serves as a
mechanism to detect if substructure finders are performing the
unbinding correctly. The unbinding errors can be masked in other
comparisons such as \vmax\ and mass plots but show up in an obvious
way when the spin distribution is examined. Phase-space finders are
less sensitive to poor unbinding as they have some implicit unbinding
in their selection criteria when looking at velocity components.
Indeed \citet{hetznecker_2006} and \citet{donghia_2007} both show
there is a good correlation between the virialisation of haloes and
the spin parameter, thus indicating its use for the determination of
how relaxed the halo is, which is not unrelated to the unbinding
process.

The mean spin parameter of subhaloes decreases as they approach the
host halo's centre. This is a real effect and not an artefact of any
difficulty in recovering structure as the subhalo approaches the
centre of the main halo. This effect is apparent in the spin parameter
distribution which matches that of field haloes at larger radii but
has a broader width than other published fits, extending to lower spin
values. 
This difference between the spin properties of subhaloes and field
haloes needs to be taken into account if precise measurement of the
spin parameter distribution are to be made.

The recovered spin parameter goes through a minimum for subhaloes near
the edge of the host at about half the \Rmax\ value. Here, if outer
particles are stripped tidally as a substructure falls into a host
halo, the result will be a decrease in the spin. This implies a radial
dependant factor needs to be taken into account when compiling
substructure catalogues, as the infalling haloes tend to have their
outer particles removed. Once the outer layer has been lost the spin
parameter generally increases to smaller radii as less and less mass
is considered.

The value of the spin parameter measured is dependent upon the
choice of where to place the outer edge and precisely which material
is included in the calculation. As we have shown here and elsewhere these
choices are very halo finder dependent and so care should be taken
when inter-comparing spin parameter measurements from different codes.

In a future project we plan to look more closely at the difference
between field and substructure haloes, to compare more directly the
spin parameter found.

\section*{Acknowledgements} \label{sec:Acknowledgements}

The work in this paper was initiated at the "Subhaloes going Notts"
workshop in Dovedale, UK, which was funded by the European
Commission’s Framework Programme 7, through the Marie Curie Initial
Training Network CosmoComp (PITN-GA-2009-238356).

We wish to thank the Virgo Consortium for allowing the use of 
the Aquarius dataset and Adrian Jenkins for assisting with the data.
The \ghalo\ datasets were kindly provided by the Zurich group.

HL and JH acknowledges a fellowship from the European Commission's Framework
Programme 7, through the Marie Curie Initial Training Network CosmoComp
(PITN-GA-2009-238356).

JH is also partially supported by NSFC 11121062, 10878001,11033006,
and by the CAS/SAFEA International Partnership Program for Creative
Research Teams (KJCX2-YW-T23).

AK is supported by the {\it Spanish Ministerio de Ciencia e
Innovaci\'on} (MICINN) in Spain through the Ramon y Cajal programme as
well as the grants AYA 2009-13875-C03-02, AYA2009-12792- C03-03,
CSD2009-00064, and CAM S2009/ESP-1496.  He further thanks La Buena
Vida for soidemersol.

PJE acknowledges financial support from the Chinese Academy of Sciences
(CAS), from NSFC grants (No. 11121062, 10878001,11033006), and by the
CAS/SAFEA International Partnership Program for Creative Research Teams
(KJCX2-YW-T23).

JO would like to thank Juli Furniss for help in revision of
this document.

\bibliography{mn-jour,SubHaloes} \bibliographystyle{mn2e}

\begin{thebibliography}{}

\bibitem[\protect\citeauthoryear{Abadi, Navarro, Steinmetz \& Eke}{Abadi
  et~al.}{2003}]{abadi_2003}
Abadi M.~G.,  Navarro J.~F.,  Steinmetz M.,    Eke V.~R.,  2003, The
  Astrophysical Journal, 591, 499

\bibitem[\protect\citeauthoryear{{Antonuccio-Delogu}, {Dobrotka}, {Becciani},
  {Cielo}, {Giocoli}, {Macci{\`o}} \& {Romeo-Velon{\'a}}}{{Antonuccio-Delogu}
  et~al.}{2010}]{antonuccio_2010}
{Antonuccio-Delogu} V.,  {Dobrotka} A.,  {Becciani} U.,  {Cielo} S.,  {Giocoli}
  C.,  {Macci{\`o}} A.~V.,    {Romeo-Velon{\'a}} A.,  2010, MNRAS, 407, 1338

\bibitem[\protect\citeauthoryear{{Ascasibar}}{{Ascasibar}}{2010}]{ascasibar_20%
10}
{Ascasibar} Y.,  2010, Comput. Phys. Commun., 181, 1438

\bibitem[\protect\citeauthoryear{Ascasibar \& Binney}{Ascasibar \&
  Binney}{2005}]{ascasibar_2005}
Ascasibar Y.,  Binney J.,  2005, MNRAS, 356, 872

\bibitem[\protect\citeauthoryear{Aubert, Pichon \& Colombi}{Aubert
  et~al.}{2004}]{adaptahop_2004}
Aubert D.,  Pichon C.,    Colombi S.,  2004, MNRAS, 352, 376

\bibitem[\protect\citeauthoryear{Barnes \& Efstathiou}{Barnes \&
  Efstathiou}{1987}]{barnes_1987}
Barnes J.,  Efstathiou G.,  1987, ApJ, 319, 575

\bibitem[\protect\citeauthoryear{{Behroozi}, {Wechsler} \& {Wu}}{{Behroozi}
  et~al.}{2011}]{behroozi_2011}
{Behroozi} P.~S.,  {Wechsler} R.~H.,    {Wu} H.-Y.,  2011, ArXiv preprint
  arXiv:1110.4372

\bibitem[\protect\citeauthoryear{Benson}{Benson}{2012}]{benson_2012}
Benson A.~J.,  2012, New Astronomy, 17, 175

\bibitem[\protect\citeauthoryear{Benson, Pearce, Frenk, Baugh \&
  Jenkins}{Benson et~al.}{2001}]{benson_2001}
Benson A.~J.,  Pearce F.~R.,  Frenk C.~S.,  Baugh C.~M.,    Jenkins A.,  2001,
  MNRAS, 320, 261

\bibitem[\protect\citeauthoryear{Bertone, De~Lucia \& Thomas}{Bertone
  et~al.}{2007}]{bertone_2007}
Bertone S.,  De~Lucia G.,    Thomas P.~A.,  2007, MNRAS, 379, 1143

\bibitem[\protect\citeauthoryear{Bett, Eke, Frenk, Jenkins, Helly \&
  Navarro}{Bett et~al.}{2007}]{bett_2007}
Bett P.,  Eke V.,  Frenk C.~S.,  Jenkins A.,  Helly J.,    Navarro J.,  2007,
  MNRAS, 376, 215

\bibitem[\protect\citeauthoryear{Bett, Eke, Frenk, Jenkins \& Okamoto}{Bett
  et~al.}{2010}]{bett_2010}
Bett P.,  Eke V.,  Frenk C.~S.,  Jenkins A.,    Okamoto T.,  2010, MNRAS, 404,
  1137

\bibitem[\protect\citeauthoryear{Bower, Benson, Malbon, Helly, Frenk, Baugh,
  Cole \& Lacey}{Bower et~al.}{2006}]{bower_2006}
Bower R.~G.,  Benson A.~J.,  Malbon R.,  Helly J.~C.,  Frenk C.~S.,  Baugh
  C.~M.,  Cole S.,    Lacey C.~G.,  2006, MNRAS, 370, 645

\bibitem[\protect\citeauthoryear{{Bryan}, {Kay}, {Duffy}, {Schaye}, {Dalla
  Vecchia} \& {Booth}}{{Bryan} et~al.}{2012}]{bryan_2012}
{Bryan} S.~E.,  {Kay} S.~T.,  {Duffy} A.~R.,  {Schaye} J.,  {Dalla Vecchia} C.,
     {Booth} C.~M.,  2012, ArXiv preprint arXiv:1207.4555

\bibitem[\protect\citeauthoryear{Bullock, Dekel, Kolatt, Kravtsov, Klypin,
  Porciani \& Primack}{Bullock et~al.}{2001}]{bullock_2001}
Bullock J.~S.,  Dekel A.,  Kolatt T.~S.,  Kravtsov A.~V.,  Klypin A.~A.,
  Porciani C.,    Primack J.~R.,  2001, ApJ, 555, 240

\bibitem[\protect\citeauthoryear{Cantalupo, Lilly \& Haehnelt}{Cantalupo
  et~al.}{2012}]{cantalupo_2012}
Cantalupo S.,  Lilly S.~J.,    Haehnelt M.~G.,  2012, MNRAS, 425, 1992

\bibitem[\protect\citeauthoryear{{Cole}, {Lacey}, {Baugh} \& {Frenk}}{{Cole}
  et~al.}{2000}]{cole_2000}
{Cole} S.,  {Lacey} C.~G.,  {Baugh} C.~M.,    {Frenk} C.~S.,  2000, MNRAS, 319,
  168

\bibitem[\protect\citeauthoryear{Croton, Springel, White, De~Lucia, Frenk, Gao,
  Jenkins, Kauffmann, Navarro \& Yoshida}{Croton et~al.}{2006}]{croton_2006}
Croton D.~J.,  Springel V.,  White S. D.~M.,  De~Lucia G.,  Frenk C.~S.,  Gao
  L.,  Jenkins A.,  Kauffmann G.,  Navarro J.~F.,    Yoshida N.,  2006, MNRAS,
  365, 11

\bibitem[\protect\citeauthoryear{Dalcanton, Spergel \& Summers}{Dalcanton
  et~al.}{1997}]{dalcanton_1997}
Dalcanton J.~J.,  Spergel D.~N.,    Summers F.~J.,  1997, ApJ, 482, 659

\bibitem[\protect\citeauthoryear{De~Lucia \& Blaizot}{De~Lucia \&
  Blaizot}{2007}]{delucia_2007}
De~Lucia G.,  Blaizot J.,  2007, MNRAS, 375, 2

\bibitem[\protect\citeauthoryear{{D'Onghia} \& {Navarro}}{{D'Onghia} \&
  {Navarro}}{2007}]{donghia_2007}
{D'Onghia} E.,  {Navarro} J.~F.,  2007, MNRAS, 380, L58

\bibitem[\protect\citeauthoryear{Eisenstein \& Hut}{Eisenstein \&
  Hut}{1998}]{hop_1998}
Eisenstein D.~J.,  Hut P.,  1998, ApJ, 498, 137

\bibitem[\protect\citeauthoryear{Elahi, Thacker \& Widrow}{Elahi
  et~al.}{2011}]{elahi_peaks_2011}
Elahi P.~J.,  Thacker R.~J.,    Widrow L.~M.,  2011, MNRAS, 418, 320

\bibitem[\protect\citeauthoryear{Fall \& Efstathiou}{Fall \&
  Efstathiou}{1980}]{fall_1980}
Fall S.,  Efstathiou G.,  1980, MNRAS, 193, 189

\bibitem[\protect\citeauthoryear{Font, Bower, McCarthy, Benson, Frenk, Helly,
  Lacey, Baugh \& Cole}{Font et~al.}{2008}]{font_2008}
Font A.~S.,  Bower R.~G.,  McCarthy I.~G.,  Benson A.~J.,  Frenk C.~S.,  Helly
  J.~C.,  Lacey C.~G.,  Baugh C.~M.,    Cole S.,  2008, MNRAS, 389, 1619

\bibitem[\protect\citeauthoryear{Frenk \& White}{Frenk \&
  White}{2012}]{frenk_2012}
Frenk C.,  White S.,  2012, Annalen der Physik, 524, 507

\bibitem[\protect\citeauthoryear{{Frenk}, {Baugh}, {Cole} \& {Lacey}}{{Frenk}
  et~al.}{1997}]{frenk_1997}
{Frenk} C.~S.,  {Baugh} C.~M.,  {Cole} S.,    {Lacey} S.,  1997, in {Persic}
  M.,  {Salucci} P.,  eds, Dark and Visible Matter in Galaxies and Cosmological
  Implications Vol.~117 of Astronomical Society of the Pacific Conference
  Series, {Numerical and Analytical Modelling of Galaxy Formation and
  Evolution}.
p.~335

\bibitem[\protect\citeauthoryear{Gill, Knebe \& Gibson}{Gill
  et~al.}{2004}]{gill_evolution_2004}
Gill S.~P.,  Knebe A.,    Gibson B.~K.,  2004, MNRAS, 351, 399

\bibitem[\protect\citeauthoryear{{Gottl{\"o}ber} \& {Yepes}}{{Gottl{\"o}ber} \&
  {Yepes}}{2007}]{gottlober_2007}
{Gottl{\"o}ber} S.,  {Yepes} G.,  2007, ApJ, 664, 117

\bibitem[\protect\citeauthoryear{Han, Jing, Wang \& Wang}{Han
  et~al.}{2011}]{han_resolving_2011}
Han J.,  Jing Y.~P.,  Wang H.,    Wang W.,  2011, Arxiv preprint
  arXiv:1103.2099

\bibitem[\protect\citeauthoryear{{Hetznecker} \& {Burkert}}{{Hetznecker} \&
  {Burkert}}{2006}]{hetznecker_2006}
{Hetznecker} H.,  {Burkert} A.,  2006, MNRAS, 370, 1905

\bibitem[\protect\citeauthoryear{Jarosik, Bennett, Dunkley, Gold, Greason,
  Halpern, Hill, Hinshaw, Kogut, Komatsu, Larson, Limon, Meyer, Nolta, Odegard,
  Page, Smith, Spergel, Tucker, Weiland, Wollack \& Wright}{Jarosik
  et~al.}{2011}]{wmap_2011}
Jarosik N.,  Bennett C.~L.,  Dunkley J.,  Gold B.,  Greason M.~R.,  Halpern M.,
   Hill R.~S.,  Hinshaw G.,  Kogut A.,  Komatsu E.,  Larson D.,  Limon M.,
  Meyer S.~S.,  Nolta M.~R.,  Odegard N.,  Page L.,  Smith K.~M.,  Spergel D.,
  Tucker G.~S.,  Weiland J.~L.,  Wollack E.,    Wright E.~L.,  2011, ApJS, 192,
  14

\bibitem[\protect\citeauthoryear{{Kauffmann}, {Nusser} \&
  {Steinmetz}}{{Kauffmann} et~al.}{1997}]{kauffmann_1997}
{Kauffmann} G.,  {Nusser} A.,    {Steinmetz} M.,  1997, MNRAS, 286, 795

\bibitem[\protect\citeauthoryear{{Kauffmann}, {White} \&
  {Guiderdoni}}{{Kauffmann} et~al.}{1993}]{kauffmann_1993}
{Kauffmann} G.,  {White} S.~D.~M.,    {Guiderdoni} B.,  1993, MNRAS, 264, 201

\bibitem[\protect\citeauthoryear{Knebe, Knollmann, Muldrew, Pearce,
  {Aragon-Calvo}, Ascasibar, Behroozi, Ceverino, Colombi \& Diemand}{Knebe
  et~al.}{2011}]{knebe_haloes_2011}
Knebe A.,  Knollmann S.~R.,  Muldrew S.~I.,  Pearce F.~R.,  {Aragon-Calvo}
  M.~A.,  Ascasibar Y.,  Behroozi P.~S.,  Ceverino D.,  Colombi S.,    Diemand
  J.,  2011, MNRAS, pp 819--

\bibitem[\protect\citeauthoryear{Knebe \& Power}{Knebe \&
  Power}{2008}]{knebe_2008}
Knebe A.,  Power C.,  2008, ApJ, 678, 621

\bibitem[\protect\citeauthoryear{Knollmann \& Knebe}{Knollmann \&
  Knebe}{2009}]{knollmann_ahf:_2009}
Knollmann S.~R.,  Knebe A.,  2009, ApJS, 182, 608

\bibitem[\protect\citeauthoryear{Lacerna \& Padilla}{Lacerna \&
  Padilla}{2012}]{lacerna_2012}
Lacerna I.,  Padilla N.,  2012, MNRAS, 426, L26

\bibitem[\protect\citeauthoryear{Macciò, Dutton, Van Den~Bosch, Moore, Potter
  \& Stadel}{Macciò et~al.}{2007}]{maccio_2007}
Macciò A.~V.,  Dutton A.~A.,  Van Den~Bosch F.~C.,  Moore B.,  Potter D.,
  Stadel J.,  2007, MNRAS, 378, 55

\bibitem[\protect\citeauthoryear{{Mestel}}{{Mestel}}{1963}]{mestel_1963}
{Mestel} L.,  1963, MNRAS, 126, 553

\bibitem[\protect\citeauthoryear{{Mo}, {Mao} \& {White}}{{Mo}
  et~al.}{1998}]{mo_1998}
{Mo} H.~J.,  {Mao} S.,    {White} S.~D.~M.,  1998, MNRAS, 295, 319

\bibitem[\protect\citeauthoryear{Muldrew, Pearce \& Power}{Muldrew
  et~al.}{2011}]{muldrew_accuracy_2011}
Muldrew S.~I.,  Pearce F.~R.,    Power C.,  2011, MNRAS, 410, 2617

\bibitem[\protect\citeauthoryear{Navarro \& Steinmetz}{Navarro \&
  Steinmetz}{2000}]{navarro_2000}
Navarro J.~F.,  Steinmetz M.,  2000, ApJ, 538, 477

\bibitem[\protect\citeauthoryear{Neyrinck, Gnedin \& Hamilton}{Neyrinck
  et~al.}{2005}]{neyrinck_voboz:_2005}
Neyrinck M.~C.,  Gnedin N.~Y.,    Hamilton A. J.~S.,  2005, MNRAS, 356, 1222

\bibitem[\protect\citeauthoryear{Onions, Knebe, Pearce, Muldrew, Lux,
  Knollmann, Ascasibar, Behroozi, Elahi, Han, Maciejewski, Merchán, Neyrinck,
  Ruiz, Sgró, Springel \& Tweed}{Onions et~al.}{2012}]{Onions_2012}
Onions J.,  Knebe A.,  Pearce F.~R.,  Muldrew S.~I.,  Lux H.,  Knollmann S.~R.,
   Ascasibar Y.,  Behroozi P.,  Elahi P.,  Han J.,  Maciejewski M.,  Merchán
  M.~E.,  Neyrinck M.,  Ruiz A.~N.,  Sgró M.~A.,  Springel V.,    Tweed D.,
  2012, MNRAS, 423, 1200

\bibitem[\protect\citeauthoryear{Peebles}{Peebles}{1969}]{peebles_1969}
Peebles P.,  1969, ApJ, 155, 393

\bibitem[\protect\citeauthoryear{Read, Hayfield \& Agertz}{Read
  et~al.}{2010}]{read_2010}
Read J.~I.,  Hayfield T.,    Agertz O.,  2010, MNRAS, 405, 1513

\bibitem[\protect\citeauthoryear{Reed, Governato, Quinn, Gardner, Stadel \&
  Lake}{Reed et~al.}{2004}]{Reed_2004}
Reed D.,  Governato F.,  Quinn T.,  Gardner J.,  Stadel J.,    Lake G.,  2004,
  MNRAS, 359, 1537

\bibitem[\protect\citeauthoryear{Sgró, Ruiz \& Merchán}{Sgró
  et~al.}{2010}]{sgro_2010}
Sgró M.~A.,  Ruiz A.~N.,    Merchán M.~E.,  2010, BAAA, 53, 43

\bibitem[\protect\citeauthoryear{Springel}{Springel}{2005}]{springel_cosmologi%
cal_2005}
Springel V.,  2005, MNRAS, 364, 1105

\bibitem[\protect\citeauthoryear{Springel, Wang, Vogelsberger, Ludlow, Jenkins,
  Helmi, Navarro, Frenk \& White}{Springel
  et~al.}{2008}]{springel_aquarius_2008}
Springel V.,  Wang J.,  Vogelsberger M.,  Ludlow A.,  Jenkins A.,  Helmi A.,
  Navarro J.~F.,  Frenk C.~S.,    White S.~D.,  2008, MNRAS, 391, 1685

\bibitem[\protect\citeauthoryear{{Springel}, {White}, {Jenkins}, {Frenk},
  {Yoshida}, {Gao}, {Navarro}, {Thacker}, {Croton}, {Helly}, {Peacock}, {Cole},
  {Thomas}, {Couchman}, {Evrard}, {Colberg} \& {Pearce}}{{Springel}
  et~al.}{2005}]{springel_millenium_2005}
{Springel} V.,  {White} S.~D.~M.,  {Jenkins} A.,  {Frenk} C.~S.,  {Yoshida} N.,
   {Gao} L.,  {Navarro} J.,  {Thacker} R.,  {Croton} D.,  {Helly} J.,
  {Peacock} J.~A.,  {Cole} S.,  {Thomas} P.,  {Couchman} H.,  {Evrard} A.,
  {Colberg} J.,    {Pearce} F.,  2005, Nat, 435, 629

\bibitem[\protect\citeauthoryear{Springel, White, Tormen \& Kauffmann}{Springel
  et~al.}{2001}]{subfind_2001}
Springel V.,  White S. D.~M.,  Tormen G.,    Kauffmann G.,  2001, MNRAS, 328,
  726

\bibitem[\protect\citeauthoryear{Stadel}{Stadel}{2001}]{stadel_2001}
Stadel J.,  2001, PhD thesis, University of Washington

\bibitem[\protect\citeauthoryear{Stadel, Potter, Moore, Diemand, Madau, Zemp,
  Kuhlen \& Quilis}{Stadel et~al.}{2009}]{Stadel_2009}
Stadel J.,  Potter D.,  Moore B.,  Diemand J.,  Madau P.,  Zemp M.,  Kuhlen M.,
     Quilis V.,  2009, MNRAS, 398, L21

\bibitem[\protect\citeauthoryear{Stadel, Wadsley \& Richardson}{Stadel
  et~al.}{2002}]{pkdgrav}
Stadel J.,  Wadsley J.,    Richardson D., , 2002, High performance
  computational astrophysics with pkdgrav/gasoline

\bibitem[\protect\citeauthoryear{{Trowland}, {Lewis} \&
  {Bland-Hawthorn}}{{Trowland} et~al.}{2012}]{trowland_2012}
{Trowland} H.~E.,  {Lewis} G.~F.,    {Bland-Hawthorn} J.,  2012, ArXiv preprint
  arXiv:1201.6108

\bibitem[\protect\citeauthoryear{{Tweed}, {Devriendt}, {Blaizot}, {Colombi} \&
  {Slyz}}{{Tweed} et~al.}{2009}]{adaptahop_2009}
{Tweed} D.,  {Devriendt} J.,  {Blaizot} J.,  {Colombi} S.,    {Slyz} A.,  2009,
  A\&A, 506, 647

\bibitem[\protect\citeauthoryear{Vitvitska, Klypin, Kravtsov, Wechsler, Primack
  \& Bullock}{Vitvitska et~al.}{2002}]{vitvitska_2002}
Vitvitska M.,  Klypin A.~A.,  Kravtsov A.~V.,  Wechsler R.~H.,  Primack J.~R.,
    Bullock J.~S.,  2002, The Astrophysical Journal, 581, 799

\bibitem[\protect\citeauthoryear{{Wang}, {Mo}, {Jing}, {Yang} \& {Wang}}{{Wang}
  et~al.}{2011}]{wang_2011b}
{Wang} H.,  {Mo} H.~J.,  {Jing} Y.~P.,  {Yang} X.,    {Wang} Y.,  2011, MNRAS,
  413, 1973

\bibitem[\protect\citeauthoryear{White \& Frenk}{White \&
  Frenk}{1991}]{white_1991}
White S.,  Frenk C.,  1991, ApJ, 379, 52

\bibitem[\protect\citeauthoryear{{White}}{{White}}{1984}]{white_1984}
{White} S.~D.~M.,  1984, ApJ, 286, 38

\bibitem[\protect\citeauthoryear{{White} \& {Rees}}{{White} \&
  {Rees}}{1978}]{white_1978}
{White} S.~D.~M.,  {Rees} M.~J.,  1978, MNRAS, 183, 341

\end{thebibliography}
\label{sec:Bibliography}

\bsp
\label{lastpage}

\end{document}